\documentclass[aps,physrev,twocolumn,groupedaddress]{revtex4-2}
\usepackage{graphicx}
\usepackage{amsmath,mathtools}
\usepackage{csquotes}
\usepackage{xcolor}
\usepackage{soul}
\sethlcolor{lightgray}
\begin{document}
\title{Cross-correlating blade--wake dynamics for a model wind turbine}
\author{Francisco J. G. de Oliveira}
\email[]{f.oliveira22@imperial.ac.uk}
\author{Zahra Sharif Khodaei}
\author{Oliver R. H. Buxton}
\affiliation{Department of Aeronautics, Imperial College London, London, UK}
\date{\today}
\begin{abstract}

Understanding how wakes interact with wind turbine blades under varying operating and inflow conditions is essential for improving fatigue prediction and performance assessment in increasingly dense wind farms. We present an experimental investigation of wake-blade coupling in a model wind turbine, focusing on the role of tip-speed ratio, $\lambda$, under varying free-stream turbulence conditions. Spatially resolved wake velocity measurements are acquired concurrently with distributed blade strain measurements using Rayleigh backscattering fibre-optic sensing, enabling direct, time-synchronised analysis of fluid-structure interaction across the blade's span. The blades' strain dynamics are strongly governed by $\lambda$, where variations of the operating condition of the turbine modify the amplitude, coherence, and the temporal/spectral organisation of the blade's structural dynamics, while free-stream turbulence primarily modulates these responses. Instantaneous joint statistics reveal negligible zero-lag dependence between wake velocity and blade strain, motivating a lagged and frequency-resolved analysis. Cycle-averaged cross-correlation and cross-power spectral density analyses demonstrate that wake-induced blade response is spatially localised within the wake shear layers and organised around rotation-coherent frequencies, with the coupling strength peaking at intermediate downstream locations. These results highlight the dominant role of operating condition in shaping wake-mediated blade loading and demonstrate the value of concurrent, spatially resolved flow-structure measurements for resolving blade-exciting flow dynamics in wind-turbine wakes. Furthermore, a consistent negative-lag peak indicates that blade strain fluctuations systematically precede downstream wake velocity fluctuations, suggesting a causal, blade-driven imprint on the wake.

\end{abstract}

\maketitle
\section{Introduction}

Wind energy has emerged as one of the fastest-growing sources of utility-scale renewable electricity, with global installed capacity now exceeding 1 TW and annual additions surpassing 100 GW in recent years \cite{WWEA2024HalfYear}. As wind farms continue to increase in size and density, a growing fraction of turbines operate under non-ideal inflow conditions, including partial or full immersion in the wakes of upstream machines. These wake interactions introduce velocity deficits, enhanced turbulence, and coherent vortex structures that alter power capture and increase unsteady loading and fatigue demands on downstream rotors \cite{Thomsen}. The aerodynamic and structural response of a turbine operating in such environments is therefore governed by a complex interplay between inflow conditions and rotor operating state. Among the parameters to monitor during wind turbine operation, tip-speed ratio-$\lambda$ (defined as $\lambda=\Omega R/U_{\infty}$ where $U_{\infty}$, $\Omega$ and $R$ correspond respectively to the free-stream velocity, and the rotor's rotational speed and radius) plays a central role in driving the turbine wake's multi-scale characteristics \cite{biswas2024a}, the nature of the blades loading linked with variations in the experienced thrust \cite{maldonado2015,thompson2023}, and power coefficient of the rotor \cite{chamorro2009,chamorro2012}, directly influencing the sectional angle of attack thereby affecting lift generation and stalling characteristics of the sectional profiles across the blade.

Moreover, $\lambda$ conditions the coherence, energy, and spectral content of wake structures shed by the rotor, including tip and root vortices \cite{vermeer2003,biswas2024a,Hansen2015Aerodynamics,gambuzza2,bourhis2025}, which in turn imprint themselves onto the blade's dynamic response. As a result, variations in $\lambda$—whether due to control actions, inflow variability, or off-design operation—can lead to substantial changes in both wake organisation and blade dynamics, highlighting the relevance of experimentally assessing how a broad range of $\lambda$ produce different wake \enquote{flavours}, how these react to different inflow \enquote{flavours} and, how the combination of the two modulates induced blade dynamics.

The interplay between free-stream turbulence (FST) and $\lambda$ further modulates the coupled wake-blade dynamics, strongly conditioned by the turbine's operating conditions ($\lambda$). Increased FST accelerates wake-recovery \cite{hodgson2025,bourhis2025,gambuzza2}, accelerates tip-vortex breakdown \cite{chamorro2009,chamorro2012,porteagel2020,biswas2025}, and alters blade boundary-layer/flow-separation behaviour \cite{maldonado2015,thompson2023}. However, the impact of FST on blade loading is not uniform across operating conditions: its influence is most pronounced at off-design $\lambda$, where stalled or transitional flow amplifies the blade's sensitivity to inflow perturbations. At high angles of attack, the presence of FST has been associated with increased lift coefficient at high angles of attack by preventing separation \cite{maldonado2015}. Moreover, FST with integral length scale of the order of the blade's chord has been associated with increased levels of fluctuating loads at fixed turbulence intensity levels in the free-stream \cite{thompson2023}. 

Fluid-structure interaction in wind turbines arises from the direct continuous exchange of momentum between the unsteady flow field generated by the rotor and the elastic response of the blades, and the indirect interaction between the elastic blades and the shed vortical structures at the rotor plane. Unlike stationary bluff bodies, turbine blades are subjected to forces that are simultaneously influenced by rotation, wake-generated coherent structures, and inflow variability, resulting in a spatially heterogeneous load distribution across their span \cite{Wen2020Blade, Trigaux2024, Bernardi2025, francisco3}. In wind farms, wind turbines are often exposed to enriched turbulence content produced by upstream turbines contributing to accumulated fatigue damage \citep{Moens2022a,Moens2022b}. In floating offshore wind turbines this is further complicated by introducing platform-relevant dynamics \citep{Zhou2025}, be it by generation of wake-enriched dynamics \citep{Messmer2025,vandenBerg2026} and their interaction with the wind turbine's body, or by modifying the local wind impacting the blade's profile. To resolve these complex interactions we require measurement strategies that resolve both the instantaneous flow conditions and instantaneous spatially resolved structural response of the turbine's blades. Such an experimental configuration provides a direct means of identifying how specific flow features—--such as shear-layer structures or rotor-synchronous vortices---are imprinted onto blade deformations. Measurements such as these are essential for disentangling aerodynamic loading from inertial contributions, and to quantify the mechanisms by which wake dynamics and operating conditions influence the fatigue-relevant blade response via cross-correlation statistical approaches \cite{zhou1999,francisco2}. 

Resolving the required level of detail experimentally requires a sensing architecture capable of resolving both the spatial distribution and temporal evolution of the blade's deformation in a way that doesn't perturb the flow development across the blade, and alter the blade's structural inertia. Conventional electrical strain gauges and accelerometers, while robust, provide only point-wise spatial measurements and hence become increasingly impractical as the number of measurement locations increases. Brillouin-based sensors or fibre Bragg grating sensors exploit the flexibility of fibre optics, with minimal intrusion into the modification of the mechanical properties of the structure and flow's deformation, are typically placed in hot-spots of structural failure \cite{Pacheco2024Experimental,Wen2020Blade,INNWIND_D224}. Such sensors can be multiplexed across a single fibre so that multiple spatial locations can be retrieved at the expense of spatially continuous information \cite{FBGsreview,FBGsmultiplex}. In contrast, Rayleigh backscattering sensors (RBS) based on optical frequency domain reflectometry (OFDR) offer high-density strain measurements within a single-mode fibre optic with fine spatial resolution ($2.6$ $\mathrm{mm}$ for a $20$ $\mathrm{m}$ long fibre optic) \cite{xu2020,francisco1,Li2025Shape}, at the compromise of temporal resolution of the measurements. This allows for blade deformation reconstruction via shape sensing algorithms \citep{xu2020,francisco2}, and to capture spatial dynamics that wouldn't be resolved otherwise. The spatial resolution allows for cross-correlations to be built between sectional regions of the blade to adjacent flow events, allowing the dynamic response of the blade to be directly correlated with wake measurements across various operating conditions. This provides a robust framework for assessing how variations in $\lambda$ and inflow turbulence jointly shape wake-induced loading.

Despite the extensive literature on the impact of inflow FST \enquote{flavours} and turbine operating conditions on separated wake dynamics and experienced wind turbine blade dynamics, experimental investigations coupling concurrently resolved wake measurements and spatially resolved blade dynamics remain scarce. While fibre-optics have been implemented in wind turbine blades in static \citep{Coscetta2017Brillouin,Tang2025FBGBlades} experiments, and under dynamically relevant operating conditions \citep{Campagnolo2013Wind,Wen2020Blade,Pacheco2024Experimental}, we perform, for the first time, an assessment of how wake-generated flow structures at different operating conditions are imprinted onto the blade's structural dynamics.

\section{Methodology}

\begin{figure}
  \raisebox{3.3in}{\textit{a)}}\includegraphics[width=\columnwidth]{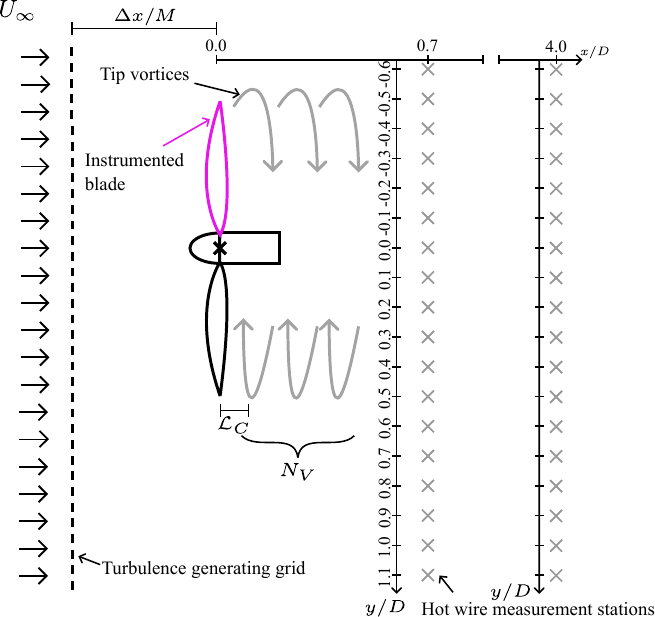}
  \raisebox{1.1in}{\textit{b)}}\includegraphics[width=\columnwidth]{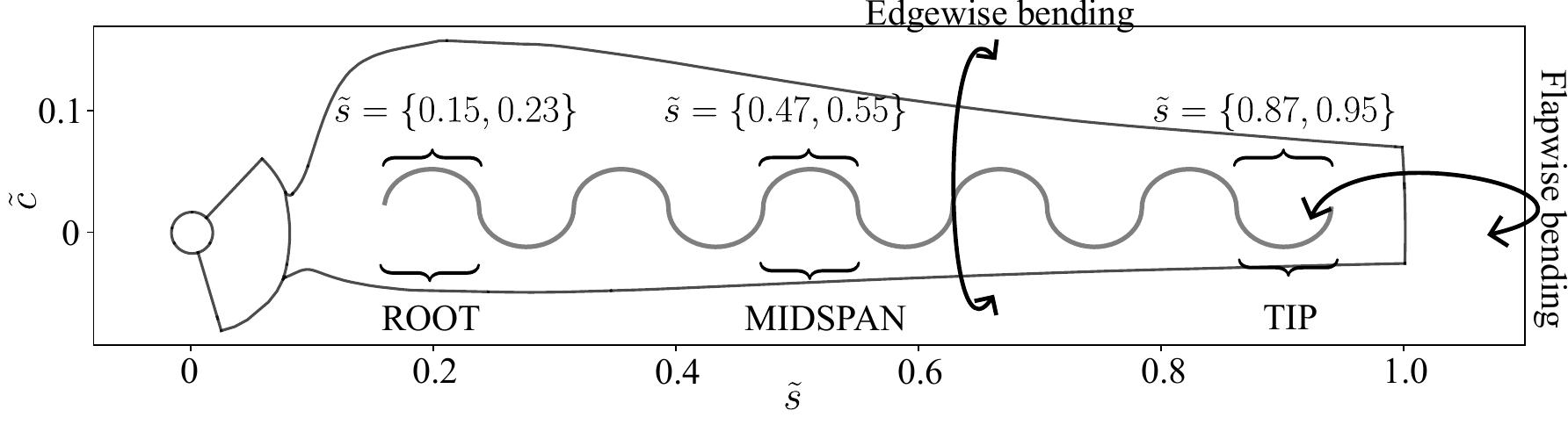}
  \caption{\textit{a)}: Experimental apparatus schematic. Depiction of the hot-wire measurement locations with respect to the turbine model. \textit{b)}: Fibre optic sensors layout across the sensorised blade. Depiction of 3 blade sections for subsequent analysis: $\{\mathrm{ROOT}, \mathrm{MIDSPAN}, \mathrm{TIP}\}$. Panel \textit{b)} is adapted from \citet{francisco3}. $\tilde{c}=c/R$ and $\tilde{s}=s/R$ correspond to  the Cartesian axis used for the blade's strain measurements.}
  \label{fig:schematic1}
\end{figure}

Experiments were carried out in the upper test section of the closed-loop $10'\times 5'$ wind tunnel in the Department of Aeronautics at Imperial College London, using the wind turbine model and measurement framework described in \cite{francisco3}. A three-bladed rotor with a diameter of $D=1\mathrm{m}$ was subjected to a constant free-stream velocity of $U_{\infty} = 2.8 \mathrm{m\,s}^{-1}$ and operated at seven distinct tip-speed ratios, $\lambda\in\{1,2,3,3.5,4,5,6.5\}$, encompassing below-design, design ($\lambda=4$), and above-design operating regimes (see \cite{francisco3} for operational details of the wind turbine model). Three distinct free-stream turbulence (FST) conditions were produced by placing a passive turbulence-generating grid upstream of the turbine at three different grid-to-turbine streamwise distances ($\Delta x / M$). This setup enabled the generation of three inflow \enquote{flavours} with distinct turbulence intensities ($\mathrm{TI}$) to which the turbine was exposed. The different inflow conditions are denoted by $\{\mathrm{A}, \mathrm{B}, \mathrm{C}\}$, characterised by $\mathrm{TI}[\%] \in \{3.8, 5.3, 8.2\}$, respectively. Figure \ref{fig:schematic1} \textit{a)} illustrates a schematic of the experimental setup.

Hot-wire anemometers were employed to characterise the turbine's near-wake at hub height, spanning positions $y/D\in\{1.1, -0.6\}$ with a spatial resolution of $\Delta y/D = 0.1$, and along the streamwise direction at stations $x/D \in\{0.7, 1, 1.5, 2, 3, 4\}$ (see figure \ref{fig:schematic1} for details). Simultaneously, distributed blade strain measurements were obtained using Rayleigh backscattering fibre optic sensors (RBS). The strain was recorded along one blade at $100 \mathrm{Hz}$, with a spatial resolution of $\delta_s=2.6 \mathrm{mm}$, covering both the spanwise and chordwise extent of the blade with the sinusoidal fibre optic path depicted in figure \ref{fig:schematic1} \textit{b)} (for more details on the fibre-optic acquisition methodology and sensor layout, see \cite{francisco3}). This configuration resulted in $240$ sensing points across the blade's span. Wake velocity fluctuations were captured with a six-probe hot-wire array sampling at $10 \mathrm{kHz}$; this data was subsequently filtered and downsampled to match the RBS acquisition rate. Both measurement systems were synchronized via a common hardware trigger. Each experimental run lasted $T=120 \mathrm{s}$, yielding $N_t \approx 1.2\times 10^4$ instantaneous strain fields with concurrent wake-velocity measurements. 
The strain can be decomposed into chordwise ($\varepsilon^{c}$) and spanwise ($\varepsilon^{s}$) components.
We have shown that the spanwise direction is the most sensitive and dynamically relevant to the flow's streamwise component and operating conditions in our previous work \cite{francisco3}. 
Similarly, \cite{Trigaux2024} have also observed increased energy in rotor operating harmonic frequencies across the spanwise component of blade dynamics when compared to the chordwise direction.
Moreover, \cite{Yadala2025JFM} have also shown in a stationary wing profile that the spanwise dynamics are the most sensitive to inflow FST. 
Hence, in this work, we focus on spanwise dynamics, which are expected to produce the largest cross-correlation with the flow's dynamics.
These can be obtained from the acquired strain signal, together with the fibre-optic sensor path's geometric layout from:
\begin{equation}
\varepsilon_{s} = \varepsilon \cdot \vert\mathrm{cos}(\theta_{f}(\tilde{s}))\vert,
\end{equation}
where $\theta_{f}$ corresponds to the local angle of the fibre path across the blade's spanwise extent, $\varepsilon$ to the measured strain, $\varepsilon_f$ to its component across the spanwise extent \cite{francisco3}, and $\tilde{s}$ to the normalised spanwise blade coordinate ($\tilde{s}=s/R$), on the Cartesian axis presented in figure \ref{fig:schematic1}. From this point onwards in this work, $\varepsilon$ refers to the spanwise component of strain.

\section{Causality considerations}\label{sec:causality}

When quantifying wake-blade coupling through cross-correlation and cross-spectral metrics, it is important to account for the impact of having spatial separation between the structural measurements in the blade, and the flow measurements downstream. As we have shown in \cite{francisco3}, and has been observed by \cite{Wen2020Blade,INNWIND_D224,Campagnolo2013Wind,Trigaux2024}, turbine blades are sensitive to selective frequencies, typically harmonics of the rotor's rotational-frequency, that are shared with wake-associated flow structures. For a three-bladed rotor operating at frequency $F_R$, the near wake, defined as the region immediately after the turbine up to $2-4$ diameters downstream \cite{crespo_survey_1999, porteagel2020, biswas2024a}, is typically populated by coherent flow structures characterised by $F_R$, $2F_R$, and $3F_R$. 
Additional coherent contributions from nacelle and tower shedding are also typically present in the near wake, while wake meandering becomes increasingly dominant further downstream \citep{biswas2024a}.
In the present configuration, the closest streamwise station at which wake velocity measurements were obtained was located at $x/D = 0.7$ downstream of the rotor plane. Consequently, the measured velocity fluctuations correspond to flow structures that have convected away from the blades prior to sampling. Among the characteristic wake frequencies, the component at $3F_R$, associated with tip-vortex shedding, introduces the largest potential bias in wake-blade causality due to its shorter convective pitch length \textit{i.e.}, the smallest streamwise distance between successive flow structures as they convect downstream. For a turbine with $N_b$ blades, the convective pitch length of successive tip vortices is given by \cite{sherry_tipvortex,biswas2024a}:

\begin{equation}
\mathcal{L}_C = \frac{\pi D (1-a)}{N_b \lambda},
\label{eq:Lc}
\end{equation}
where $a$ corresponds to the axial induction factor of the wind turbine. The corresponding number of vortices separating the blade from a wake measurement location at streamwise distance $X$ is:
\begin{equation}
	N_V = \frac{N_b X \lambda}{\pi D (1-a)} = \frac{X}{\mathcal{L}_C}.
\end{equation}

\begin{figure}
  \includegraphics[width=\columnwidth]{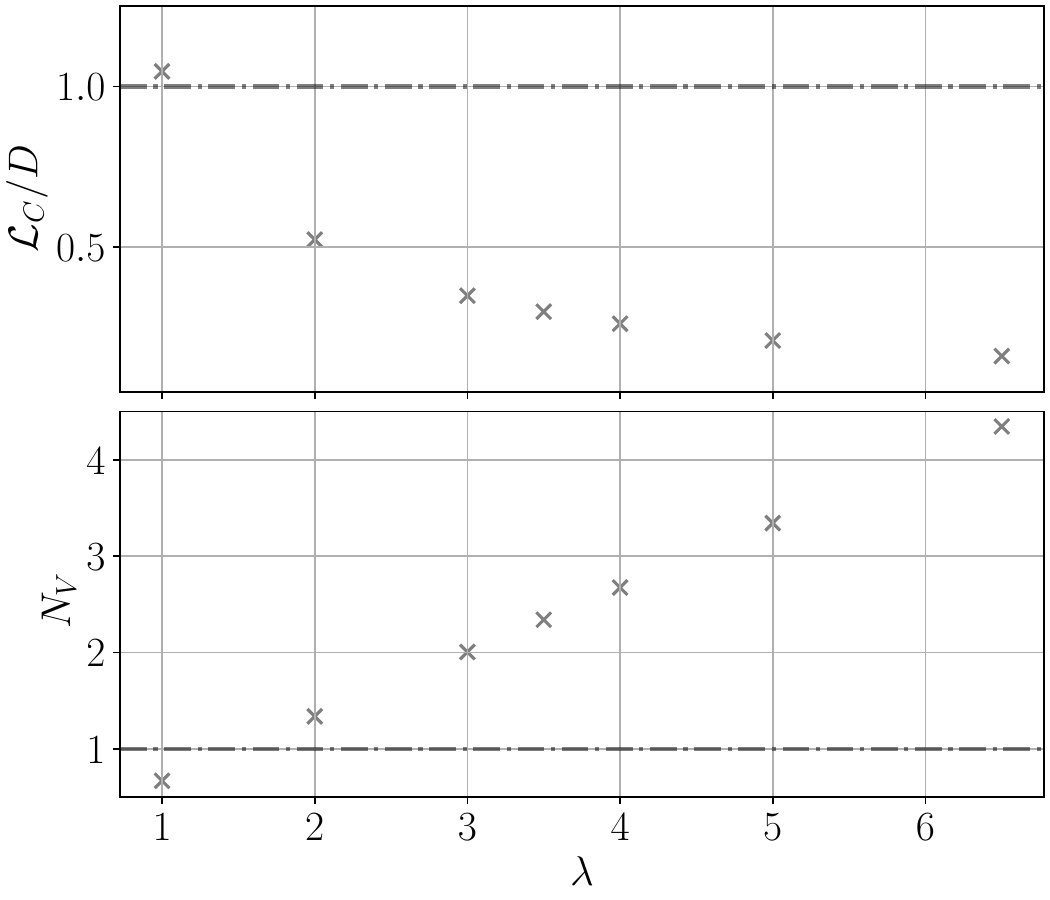}
  \caption{Convection length scale of tip-generated flow structures, and number of flow structures between the first streamwise hot-wire measurement station (at $x/D = 0.7$) and the blade, for the tested $\lambda$. The dark horizontal dashed-line on the bottom panel corresponds to the lower limit at which we expect to have no bias in causality ($N_V < 1$).}
  \label{fig:LcNv}
\end{figure}

The parameter $N_V$ therefore provides a compact, non-dimensional measure of the bias introduced by vortex advection between the strain and velocity measurements. As $N_V$ increases, the wake signal increasingly reflects the cumulative downstream evolution of multiple vortical structures rather than the blade-localised forcing associated with their generation. This effect is expected to manifest as an increasing temporal offset and reduced direct correspondence in wake-blade correlations as $\lambda$ increases. To obtain a lower-bound estimate of the convective pitch length, we assume $a \approx 0$---the same as assuming that tip-vortices are advected by $U_{\infty}$ instead of $U_{\infty}(1-a)$---providing a \enquote{best case scenario} where $N_V$ is minimized, presenting a minimum estimate of vortex-induced decorrelation effects. Figure \ref{fig:LcNv} presents the variation of the convective pitch length $\mathcal{L}_C$ (top panel) and the number of intervening vortices $N_V$ (bottom panel) as a function of $\lambda$. 

As $\lambda$ increases, the convective pitch length of the tip vortices decreases, leading to a monotonic increase in $N_V$. Notably, only for $\lambda = 1$ does the first wake measurement station present $N_V < 1$, implying minimal bias between the blade-generated vortical structures and their measured wake signatures. As $N_V$ increases, the measured wake signal reflects the superposition of the downstream evolution of multiple vortical structures rather than the one-to-one correspondence between an acting flow structure to the respective blade forcing, introducing an inherent bias in causal interpretation based solely on instantaneous time lags between the two signals. Therefore, the present analysis quantifies wake-mediated, rather than near-field aerodynamic, contributions to blade dynamics. Considering $a>0$ would increase $N_V$ and decrease $\mathcal{L}_C$, strengthening the arguments made here.

\section{Dynamical response of the blade}

We start the analysis by assessing how $\lambda$ and FST conditions impact the blade's dynamical structural response. We define $\tilde{x}=x/D$, $\tilde{y}=y/D$ to simplify notation. The measured strain signal ($\varepsilon_{\mathrm{m}}(\tilde{s},t)$) is first decomposed into its fluctuating counterpart ($\varepsilon(\tilde{s},t)$):
\begin{equation}
	\varepsilon_{\mathrm{m}}(\tilde{s},t) = \overline{\varepsilon(\tilde{s})} + \varepsilon(\tilde{s},t),
\end{equation}
where $\overline{\varepsilon(\tilde{s})}$ corresponds to the time-averaged strain at location $\tilde{s}$. Similarly, the measured velocity signal at station $(\tilde{x},\tilde{y})$ is decomposed into its fluctuating counterpart, $u_{\mathrm{m}}(\tilde{y},\tilde{x})=\overline{u(\tilde{y},\tilde{x})}+u(\tilde{y},\tilde{x})$, where $u_{\mathrm{m}}$ and $\overline{u}$ correspond respectively to the measured, and time-averaged velocities.

To characterise the structure of the blade's temporal response and its dominant time scales, we compute the autocorrelation function of $\varepsilon(\tilde{s},t)$, defined as:
\begin{equation}
	\mathrm{R}^{\varepsilon^{\ast}}_{i}(\tau) = \langle \varepsilon^{\ast}(\tilde{s}_i, t) \varepsilon^{\ast}(\tilde{s}_i, t+\tau) \rangle,
\label{eq:AUTOCORR}
\end{equation}
where $\langle \cdot \rangle$ denotes time averaging over the duration of each run, and $(\cdot)^{\ast}$ to $\varepsilon$ normalised by its standard deviation, $\sigma_{\varepsilon}$ ($\varepsilon^{\ast}=\varepsilon/\sigma_\varepsilon$). To obtain a representative measure of the structural response at different blade regions, the autocorrelation functions are averaged across sensing points within the blade sections defined in figure \ref{fig:schematic1} \textit{b)} [$\mathrm{ROOT}$, $\mathrm{MIDSPAN}$, $\mathrm{TIP}$]:
\begin{equation}
\mathrm{R}^{\circ}_{\varepsilon^{\ast}}(\tau) = \frac{1}{N_s}\sum_{i=1}^{N_s} R^{\varepsilon^{\ast}}_{i}(\tau),
\label{eq:AUTOCORRAVG}
\end{equation}
where $N_s$ is the number of sensing locations within each section. The time lag $\tau$ is normalised by the rotor period, such that $\tau^\star = \tau F_R$ for each $\lambda$. The autocorrelation function quantifies the temporal coherence of strain fluctuations: high values at lag $n \tau$ where $n$ is an integer, indicate that strain fluctuations exhibit temporal periodicity with characteristic time scale $\tau$, while rapid decay suggests predominantly stochastic, uncorrelated dynamics. By averaging across sensing points within each blade section, $R^{\circ}_{\varepsilon^{\ast}}(\tau)$ provides a representative measure of the dominant temporal scales governing the structural response in that region.

\begin{figure*}
  \includegraphics[width=\textwidth]{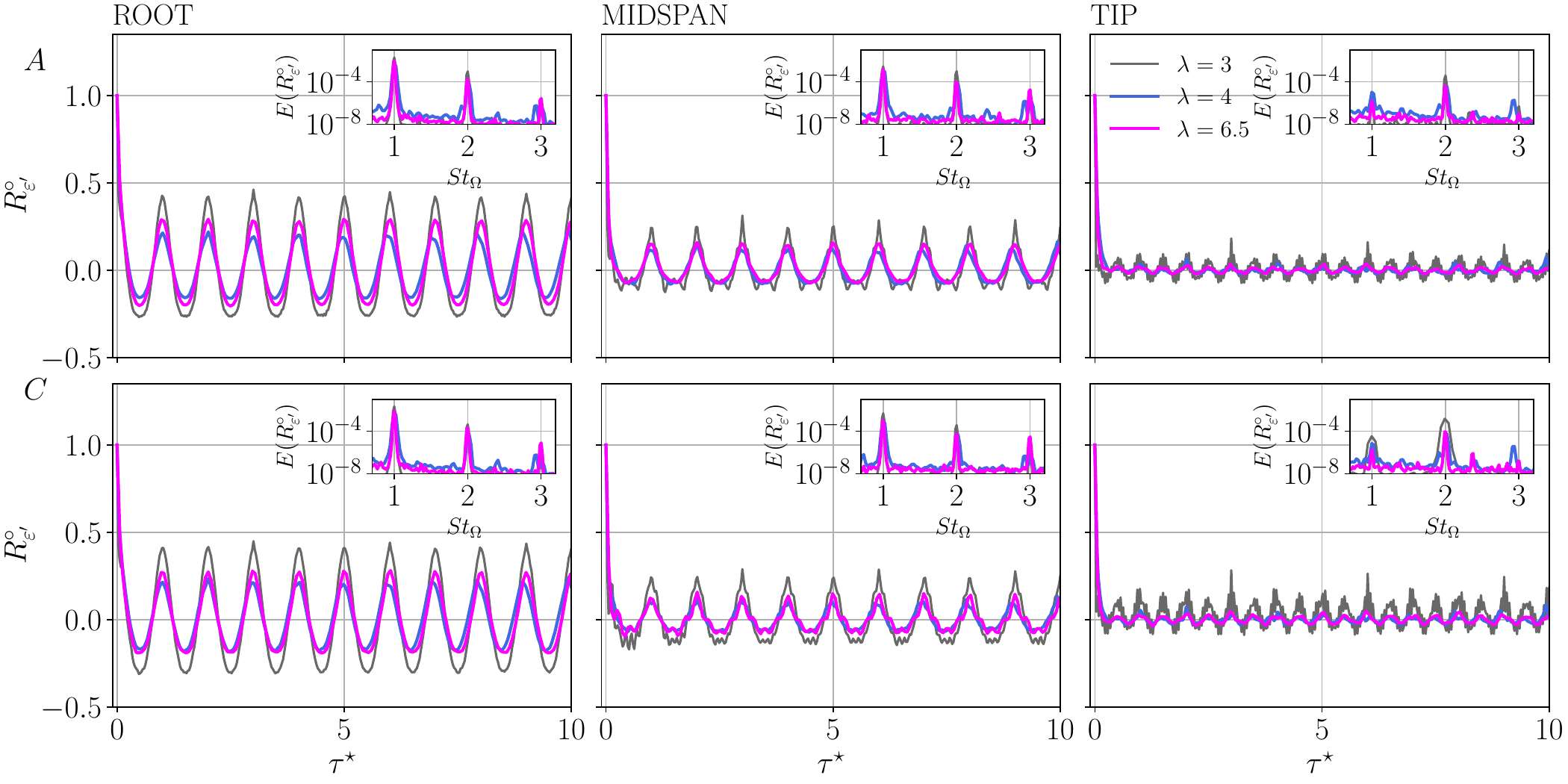}
  \caption{Autocorrelation function $R_{\varepsilon^{\ast}}^{\circ}$ for FST conditions $\mathrm{A,C}$, $\lambda\in\{3,4,6.5\}$ at the $\{\mathrm{ROOT},\mathrm{MIDSPAN}, \mathrm{TIP}\}$ blade sections. The insets correspond to the power spectral density distributions of the corresponding $R_{\varepsilon^{\ast}}^{\circ}$.}
  \label{fig:AUTOCORR}
\end{figure*}

Figure \ref{fig:AUTOCORR} shows the blade-section-averaged autocorrelation functions $R_{\varepsilon^{\ast}}^{\circ}$ for FST conditions $\mathrm{A}$ and $\mathrm{C}$, as well as selected tip-speed ratios $\lambda \in \{3,4,6.5\}$, evaluated at the $\mathrm{ROOT}$, $\mathrm{MIDSPAN}$, and $\mathrm{TIP}$ blade sections. In all cases, the autocorrelation functions display a distinct structure, reflecting the influence of strong periodic components in the strain response. From the $\mathrm{ROOT}$ to the $\mathrm{TIP}$, the second peak of the autocorrelation function diminishes in magnitude, suggesting an increase in predominantly stochastic behaviour, with increasingly short temporal coherence. Increasing the FST $\mathrm{TI}$ from condition $\mathrm{A}$ to $\mathrm{C}$ slightly increases the amplitude of the second peak of $R_{\varepsilon^{\ast}}^{\circ}$, and the amplitude of its oscillations over $0$, especially observed at the $\mathrm{TIP}$. This trend is consistent with more pronounced stochastic forcing and a diminished presence of periodic load signatures, likely due to an earlier breakdown of tip vortices and other coherent flow structures in the wake of a wind turbine operating under high $\mathrm{TI}$ \cite{biswas2025,bourhis2025}.

The insets in figure \ref{fig:AUTOCORR} present the power spectral density (PSD) of the corresponding depicted autocorrelations ($E(R_{\varepsilon^{\ast}}^{\circ})$) as a function of $St_{\Omega}$, the frequency normalised by the turbine's rotating frequency ($St_{\Omega}=f/F_R$). The PSD of $R_{\varepsilon^{\ast}}^{\circ}$ is directly linked with the PSD of $\varepsilon$ at each blade's section, allowing us to infer the blade's strain signal's-spectra as, according to the Wiener-Khinchin theorem \cite{Wienchen}, considering a temporal signal $\xi(t)$, and its autocorrelation function $R_{\xi\xi}$:
\begin{equation}
  \vert\mathcal{F}(\xi(t))\vert^2 = \mathcal{F}(R_{\xi \xi}(\tau)),
\end{equation}
where $\mathcal{F}$ corresponds to the Fourier operation term. The insets of figure \ref{fig:AUTOCORR} further corroborate the influence of periodic load components on the blade's strain signal, showing that all three sections of the blade under analysis are dominated by discrete peaks at harmonics of the rotor frequency, similarly to what has been observed by \citep{Wen2020Blade,INNWIND_D224,Moens2022a,Moens2022b,Trigaux2024}. The energy associated with $St_{\Omega}=1$ is the strongest at the $\mathrm{ROOT}$, decreasing as we move towards the $\mathrm{TIP}$. The $\mathrm{ROOT}$ and the $\mathrm{MIDSPAN}$ are mainly  driven by $St_{\Omega}\in\{1,2\}$ associated dynamics, while the $\mathrm{TIP}$ is mainly influenced by $St_{\Omega}=2$. The influence of $St_{\Omega}=3$ driven dynamics is clear in all $3$ sections, with smaller relative contributions at the $\mathrm{ROOT}$ and $\mathrm{MIDSPAN}$, and larger at the $\mathrm{TIP}$. Furthermore, the contribution of $St_{\Omega}=3$ increases non-monotonically with $\lambda$ at the $\mathrm{TIP}$, peaking at design operating conditions ($\lambda=4$) as we previously observed in \cite{francisco3}. These frequency-selective responses motivate the subsequent frequency-resolved cross-spectral analysis, which isolates the wake-mediated contributions at each harmonic. 

\begin{figure*}
  \includegraphics[width=\textwidth]{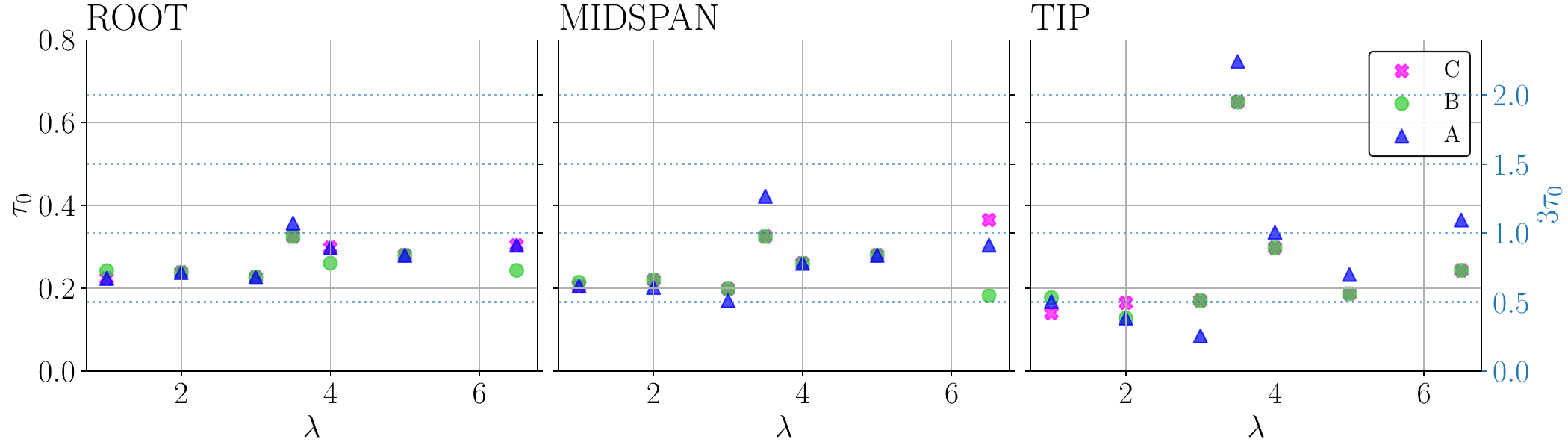}
  \caption{First zero-crossing of $R_{\varepsilon^{\ast}}^{\circ}$ for the full envelope of $\lambda$ and FST cases tested at the $\{\mathrm{ROOT},\mathrm{MIDSPAN}, \mathrm{TIP}\}$ blade sections.}
  \label{fig:TAU0}
\end{figure*}

To provide a quantitative measure of the temporal coherence of the mechanical response of the blade, figure \ref{fig:TAU0} presents the first zero-crossing time-lag ($\tau_0$) of $R_{\varepsilon^{\ast}}^{\circ}$ as a function of $\lambda$ for all FST conditions. $\tau_0$ is presented in a two $y$-inset axis, the left $y$-axis is normalised with the rotor-characteristic period as we did with $\tau^{\star}$---where $\tau_0=1$ corresponds to the period of a full rotor rotation--- the right-axis corresponds to the rescaling of $\tau_0$ with the tip-vortex-characteristic period, where $3\tau_0=1$ corresponds to the tip-vortex characteristic period. Across the operating envelope, $\tau_0$ remains of order $\mathcal{O}(0.2\text{--}0.4)$ rotor periods at the $\mathrm{ROOT}$ and $\mathrm{MIDSPAN}$, indicating that for these blade sections, the dominant coherent contributions to the strain response occur on sub-revolution time scales associated with operation-related and wake-induced loading.

For all three blade sections, $\tau_0$ reaches a local maximum at $\lambda = 3.5$, following a similar evolution to that observed for the strain fluctuation intensity as a function of $\lambda$ as we previously observed in \cite{francisco3}. This operating condition, located just below design operating conditions, is characterised by enhanced unsteadiness in the strain response, consistent with intermittent reattachment–detachment dynamics associated with the onset of blade stall. Under these operating conditions ($\lambda=3.5$), the spanwise evolution across blade sections of $\tau_0$ significantly increases as we progress from the $\mathrm{ROOT}$ towards the $\mathrm{TIP}$. When the first zero-crossing is expressed in units of the tip-vortex passing period (right axis, where $3\tau_0 = 1$), the $\mathrm{ROOT}$ and $\mathrm{MIDSPAN}$ sections exhibit $3\tau_0\approx 1$ at $\lambda=3.5$. This indicates that the strain coherence time scale coincides with the characteristic period between successive blade passages, linking the temporal structure of blade loading to the formation and advection of tip vortices in the near wake.   

At design operating conditions ($\lambda = 4$), $3\tau_0$ increases from the $\mathrm{ROOT}$ towards the $\mathrm{TIP}$, where $3\tau_0\approx 1$. This spanwise evolution suggests a progressive alignment of the strain coherence time scale with the characteristic tip-vortex period towards the tip of the blade at design operating conditions. Such behaviour is consistent with the enhanced energy associated with $3F_R$ structural dynamics at the $\mathrm{TIP}$ under design operating conditions \cite{francisco3}. For $\lambda < 3.5$, the $\mathrm{TIP}$ consistently exhibits smaller values of $\tau_0$ than the $\mathrm{ROOT}$ and $\mathrm{MIDSPAN}$, indicating shorter correlation time scales associated with shorter time-spans of coherent loading mechanisms acting on the blade. For $\lambda>4$, the spanwise evolution of $\tau_0$ at fixed $\lambda$ presents decreased spanwise scatter across blade sections when compared to $\lambda<3.5$, suggesting an increased homogeneity in the distribution of loads across the blade. The influence of FST is secondary to that of $\lambda$. For $\lambda \geq 3.5$, the FST condition $\mathrm{A}$ is associated with comparatively larger values of $\tau_0$ across blade sections. In contrast, for $\lambda\leq3$ the same inflow condition yields the smallest values of $\tau_0$, indicating that the influence of FST on the coherence of blade loading depends more on the operating regime.
Moreover, the absence of $\tau_0 \approx 1$ across all cases suggests that the fundamental rotor frequency $F_R$ is not the dominant timescale governing the blade's response. Instead, most results fall within the range $0.2 \lesssim \tau_0 \lesssim 0.4$, indicating dynamics occurring on faster timescales than those associated with $F_R$ and $2F_R$. This points to a sensitivity to higher-frequency structures that are either directly or indirectly linked to $3F_R$ and tip-vortices, particularly near $\lambda \approx 4$.

The autocorrelation analysis establishes that blade strain dynamics are dominated by rotation-coherent frequencies ($St_\Omega \in \{1,2,3\}$) with coherence times generally of order $0.2-0.4$ rotor periods. In the following sections, we investigate how these temporal structures arise from wake-blade coupling by cross-correlating blade strain with concurrent wake velocity measurements.

\section{Spatio-temporal cross-correlation of wake-blade dynamics}

The concurrent acquisition of wake velocity fluctuations and distributed blade strain dynamics allows for a direct, time-resolved investigation of wake-blade coupling. We can then quantify the temporal relationship between velocity fluctuations in the wake, and induced strain response in the different blade sections, without relying on indirect or statistical alignment of independent datasets.

We begin by assessing the joint probability density function ($\mathrm{jPDF}$) of spatially aggregated wake velocity and blade strain fluctuations, to determine whether any instantaneous statistical dependence exists between the concurrently acquired signals. To obtain a representative strain time series, we sum the fluctuating strain $\varepsilon(\tilde{s},t)$ along the blade's spanwise extent, from the first to the last measurement station (denoted by, $\tilde{s}_i$ to $\tilde{s}_e$). Similarly, a single wake-velocity time series is constructed by summing the fluctuating wake velocity $u(\tilde{x}=0.7,\tilde{y},t)$ across all transverse wake measurement locations at the first streamwise measurement station ($\tilde{x}=0.7$), from the initial to the final spanwise measurement station (i.e., $\tilde{y}_i$ to $\tilde{y}_e$). We focus on measurements taken at $\tilde{x}=0.7$ to reduce the causal bias induced by flow advection previously discussed. The resulting spatially aggregated signals, denoted by the subscript $\zeta$, are then normalized by their respective temporal standard deviations ($\sigma_{\zeta}$), yielding:
\begin{equation}
	\varepsilon_{\zeta}^{\ast}(t) = \frac{\sum_{s=\tilde{s}_i}^{\tilde{s}_e}\varepsilon(s,t)}{\sigma_{\varepsilon_{\zeta}}}, u^{\ast}_\zeta(\tilde{x},t) = \frac{\sum_{y=\tilde{y}_i}^{\tilde{y}_e} u(\tilde{x},\tilde{y},t)}{\sigma_{u_{\zeta}}}.
\end{equation}

The joint probability density function $f(\varepsilon_{\zeta}^{\ast},u_{\zeta}^{\ast})$ is then computed from the distributions of $\varepsilon_{\zeta}^{\ast}(t)$ and $u_{\zeta}^{\ast}(\tilde{x},t)$.

\begin{figure*}
  \includegraphics[width=\textwidth]{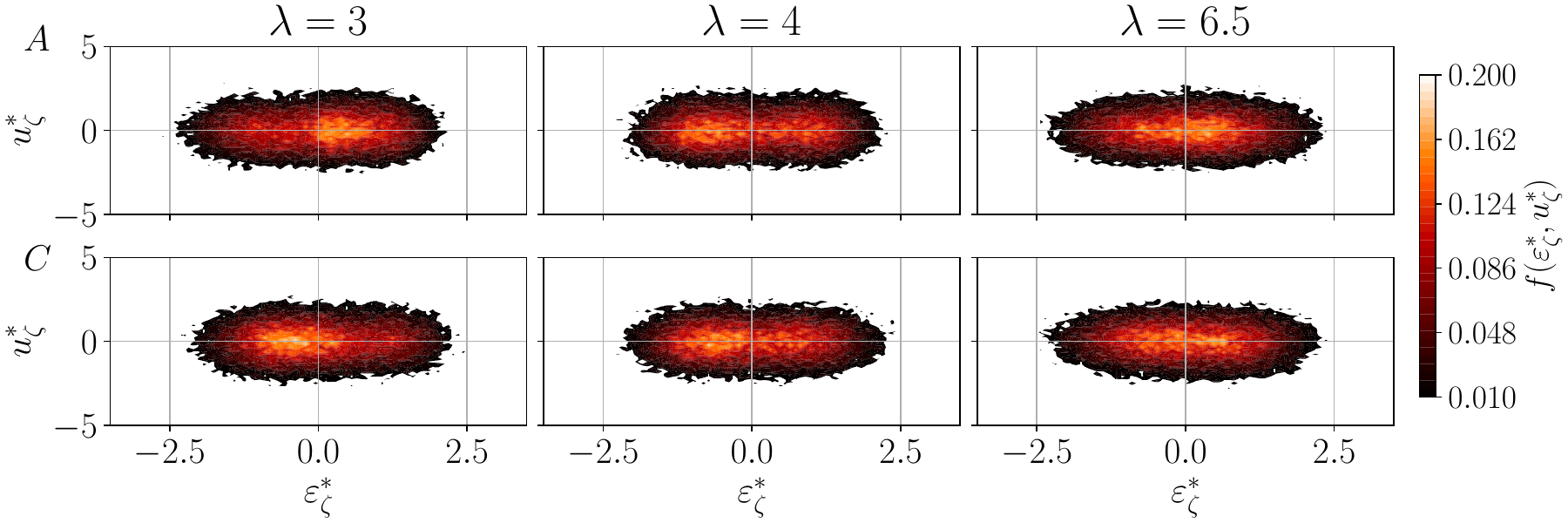}
  \caption{Joint-probability density function ($f(\varepsilon_{\zeta}^{\ast},u_{\zeta}^{\ast})$) between the blade's strain and near-wake velocity fluctuations  at $\tilde{x} = 0.7$.}
  \label{fig:JPDF}
\end{figure*}

Figure \ref{fig:JPDF} presents $f(\varepsilon_{\zeta}^{\ast},u_{\zeta}^{\ast})$ for representative $\lambda$ and FST conditions at $\tilde{x}=0.7$. Across all cases, the $\mathrm{jPDFs}$ remain approximately axis-aligned exhibiting weak variations in shape. The absence of a pronounced tilt or elongation of the distributions indicates negligible instantaneous linear dependence between the aggregated wake and blade's strain response. This result is robust across operating and FST conditions, suggesting that zero-lag joint statistics do not capture the dominant wake-blade interaction mechanisms in the present configuration. This lack of structure in the instantaneous $\mathrm{jPDFs}$ does not imply an absence of wake-blade coupling, rather reflecting the inherently time-delayed and phase-dependent nature of the interaction under analysis. To resolve this delayed coupling, lagged and frequency-resolved metrics such as cross-correlations and cross-power spectral correlations are called upon.

Cross-correlations between $u$ and $\varepsilon$ allow us to retrieve coupling dynamics between the wake-blade system by accounting for time-lags between the two signals. The cross-correlation function between the two signals, normalised by their corresponding standard deviation, is defined as:
\begin{equation}
	R_{u^{\ast}\varepsilon^{\ast}}(\tilde{x}, \tilde{y}, \tilde{s},\tau) =	\frac{1}{T}\int_{0}^{T} u^{\ast}(\tilde{x},\tilde{y},t) \varepsilon^{\ast}(\tilde{s},t+\tau) \mathrm{d}t,
\label{eq:correlation}
\end{equation}
where $\tau$ is the time lag and $T$ is the finite measurement duration ($T = 120\mathrm{s}$). This provides a global measure of the temporal correlation between wake dynamics at a given wake spatial location, and the blade's strain response at a specified blade location, accounting for both positive and negative time lags. Furthermore, by cross-correlating the concurrently measured blade's strain, and the wake's velocity fluctuations, we isolate the aerodynamically driven dynamics present in the wake-measurement station from the total strain dynamics encompassing both the aerodynamic-driven loads, and the quasi-steady centrifugal loads. The latter are uncorrelated with the unsteady flow field, and are expected not to contribute to cross-correlation statistics henceforth described. 

The time lag $\tau$ between the two concurrent signals that maximises the magnitude of $R_{u^{\ast}\varepsilon^{\ast}}(\tilde{x}, \tilde{y}, \tilde{s},\tau)$ gives the indication of the time difference between a measured velocity fluctuation and its imprint onto the blade's response. However, as previously reported, the time-lags associated with $R_{u^{\ast}\varepsilon^{\ast}}(\tilde{x}, \tilde{y}, \tilde{s},\tau)$ will inherently be populated with the bias in causality between the measured flow structures at streamwise station $\tilde{x}$, and their effect on the blade's structural response. To mitigate the resulting bias, we compute a cycle-averaged cross-correlation over a revolution period defined as:
\begin{multline}
	R_{u^{\ast}\varepsilon^{\ast}}^{\star}(\tilde{x}, \tilde{y}, \tilde{s},\tau) = \frac{1}{N_{\mathrm{cycles}}}\\
	\sum_{n=1}^{N_{\mathrm{cycles}}} \frac{1}{t_{\Omega}} \int_{0}^{t_{\Omega}} u_{n}^{\ast}(\tilde{x},\tilde{y},t + n t_{\Omega}) \varepsilon_{n}^{\ast}(\tilde{s},t+n t_{\Omega} + \tau) \mathrm{d}t,
\label{eq:correlation_star}
\end{multline}
where $N_{\mathrm{cycles}}$ is the number of rotor cycles within the total acquisition time $T$, $n$ is an integer and $t_{\Omega}$ is the period of each cycle. Averaging cross-correlations over complete rotor cycles emphasizes repeatable, rotation-synchronized coupling patterns. This allows for meaningful comparison of coupling strength across operating conditions, though the spatial offset between measurement locations---and the associated convective time delay---remains inherent to all measurements.

\begin{figure*}
  \centering
  \includegraphics[width=\textwidth]{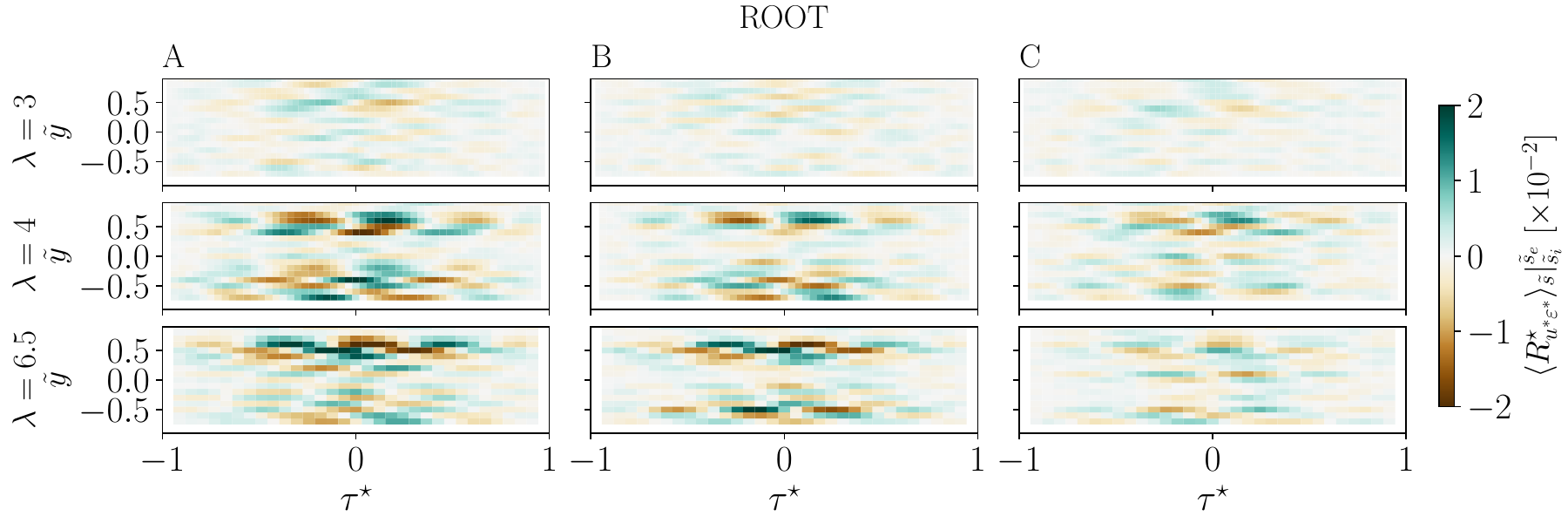}  
  \includegraphics[width=\textwidth]{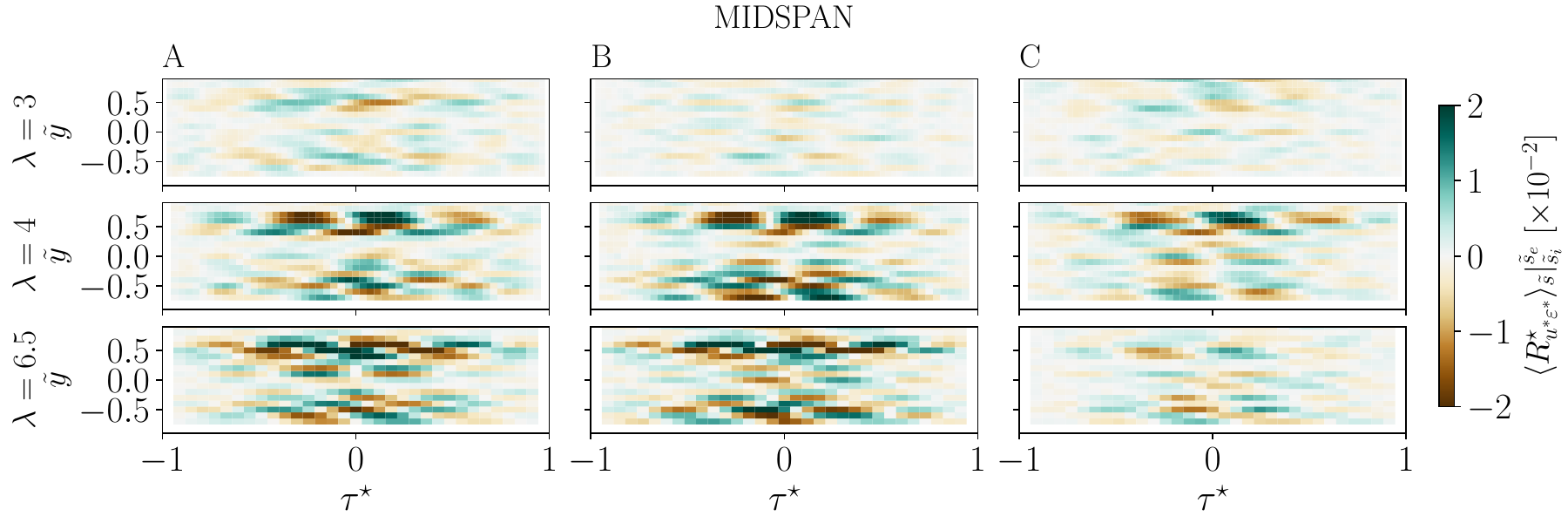}
  \includegraphics[width=\textwidth]{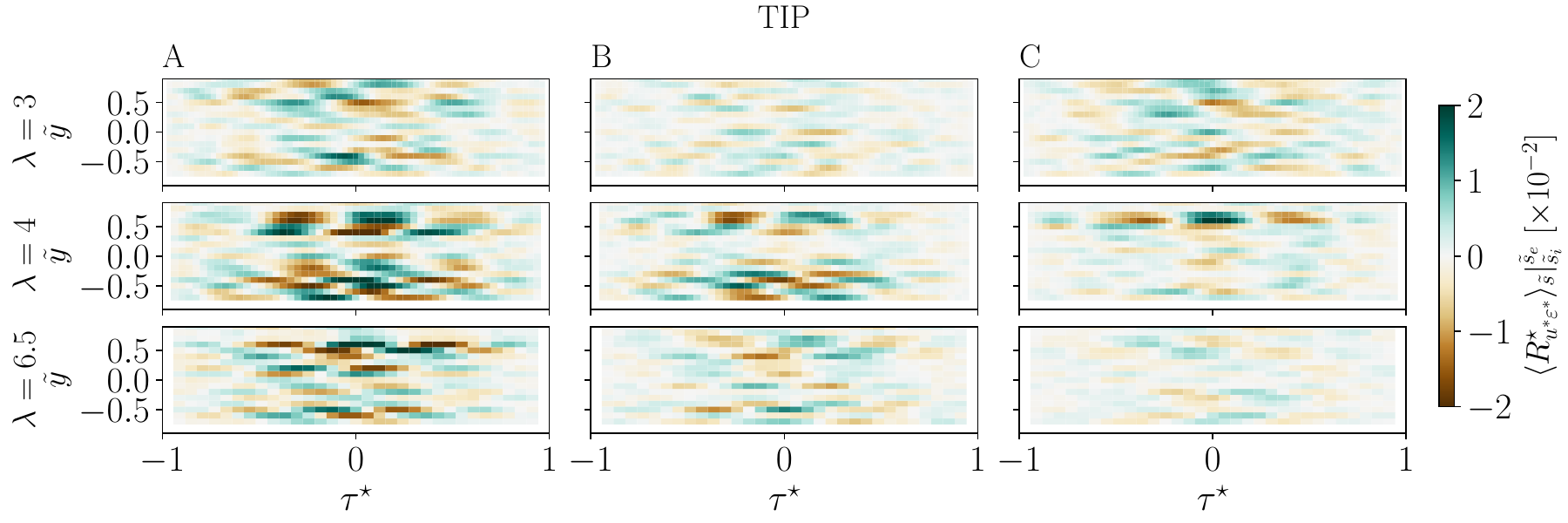}
  \caption{Cycle-averaged cross-correlation $\langle R^{\star}_{u^{\ast}\varepsilon^{\ast}}(\tilde{y},\tau^\star)\rangle$ between $u^{\ast}$ and $\varepsilon^{\ast}$ at $\mathrm{ROOT}$, $\mathrm{MIDSPAN}$, and $\mathrm{TIP}$, for the three tested inflow conditions $\mathrm{A-C}$ and $\lambda \in \{3,4,6.5\}$, as a function of $\tilde{y}$ and $\tau^\star$, at $\tilde{x}=0.7$.}
  \label{fig:fig13new}
\end{figure*}

Figure \ref{fig:fig13new} presents $R^{\star}_{u^{\ast}\varepsilon^{\ast}}$, between velocity fluctuations acquired at $\tilde{x}=0.7$ (the first streamwise measurement station) and the blade's strain fluctuations as a function of $\tilde{y}$, $\tau^\star$ (where the time lag has been normalised by the revolution period of the rotor), for the $3$ blade sections under analysis, representative $\lambda \in \{3,4,6.5\}$ and FST conditions tested. The correlations are shown at $\tilde{x}=0.7$, the closest wake measurement station downstream of the rotor, reflecting the combined effects of the streamwise development of blade-generated flow structures, and their interaction with the blade's structural response. For each of the presented maps of $R_{u^{\ast}\varepsilon^{\ast}}^{\star}(\tilde{x}, \tilde{y}, \tilde{s},\tau)$, the cross-correlations have been ensemble-averaged over the measurement locations spanning each section of the blade---$\langle R^{\star}_{u^{\ast}\varepsilon^{\ast}}\rangle_{\tilde{s}}\vert_{s_i}^{s_e}$, where $s_i$ and $s_e$ correspond to the spatial limits delimiting each region.

As seen in figure \ref{fig:fig13new} the dominant correlation signatures are consistently located within the wake's shear layers $\tilde{y}\approx \pm 0.5$ across all operating conditions. The wake's core region ($\tilde{y} \in [-0.3,0.3]$) exhibits comparatively weaker correlation, indicating that the blade's dynamic response to flow structures in the wake is primarily shear-layer driven. Out of the $3$ sections under analysis, the $\mathrm{MIDSPAN}$ presents the largest correlation with wake-core dynamics.

The cross-correlation maps show a clear dependence on blade section and $\lambda$. $\lambda=3$ presents the smallest magnitude of $\langle R^{\star}_{u^{\ast}\varepsilon^{\ast}}(\tilde{y},\tau^\star)\rangle$ out of the selected $\lambda$, for all FST cases and blade sections. Meanwhile, $\lambda\in\{4,6.5\}$ presents a significant increase of $\langle R^{\star}_{u^{\ast}\varepsilon^{\ast}}(\tilde{y},\tau^\star)\rangle$. This variation is linked with the shift in operation from below-to-above design conditions, where the blade's response is postulated to be increasingly sensitive to wake dynamics \cite{francisco3}. 
At below-design $\lambda$, the flow is separated across the blade and the wake is populated with \enquote{stochastic}, incoherent velocity fluctuations. As $\lambda$ increases towards the design $\lambda$, coherent and more energetic flow structures span the wake, potentially feeding back onto the mechanical response of the wind turbine's blades. As $\lambda$ increases past design operating conditions, the energy of these flow structures decreases followed up by a decrease in the respective blade fluctuating dynamics \citep{francisco3}.
The cross-correlation maps show consistently an increased magnitude of $\langle R^{\star}_{u^{\ast}\varepsilon^{\ast}}(\tilde{y},\tau^\star)\rangle$ in $\tilde{y}=0.5$, possibly due to a small yaw misalignment between the rotor's plane and the free-stream, seen to significantly modify the wake evolution throughout the streamwise extent of the flow \cite{vandenbroek2023yaw,bossuyt2021wake,bastankhah2016yaw}. The $\mathrm{ROOT}$, shows the smallest correlation to the turbine's wake, whilst the $\mathrm{MIDSPAN}$ emerges as the blade section that presents the largest correlation to the measured wake dynamics. Towards the $\mathrm{TIP}$, the correlations become increasingly spatially diffuse, while maintaining the spatial structure previously described, peaking in magnitude at $\lambda=4$.

The increase of FST $\mathrm{TI}$ affects the magnitude and spatial organisation of the shear-layer-driven correlations rather than their location. As $\mathrm{TI}$ increases, the correlation magnitude significantly decreases. Moreover, the correlation maps become progressively more diffuse across $\tilde{y}$, with FST case $\mathrm{A}$ presenting clear peaks in the shear-layer region of the wake.

The presence of FST has been associated with increased wake mixing, and with an earlier decay/breakdown of coherent flow structures developing in the rotor's near-wake \cite{bourhis2025,biswas2025,chamorro2009}. We observe that the presence of increased FST $\mathrm{TI}$ promotes decorrelation of the wake-blade coupling, attributed to the early onset of breakdown of the flow mechanisms driving the structural dynamics in the blade.

The cycle-averaged cross-correlation $R^{\star}_{u^{\ast}\varepsilon^{\ast}}$ quantifies coupling strength but, as discussed in Section \ref{sec:causality}, cannot reliably establish causality when $N_V > 1$. To identify leading/lagging dynamics between wake velocity and blade strain, we extend the cycle-averaging window to encompass 15 revolution periods (statistical convergence was observed after 10 revolution periods):

\begin{multline}
	R_{u^{\ast}\varepsilon^{\ast}}^{\star\star}(\tilde{x}, \tilde{y}, \tilde{s},\tau) = \frac{1}{N_{15,\mathrm{cycles}}}\\
	\sum_{n=1}^{N^{15}_{\mathrm{cycles}}} \frac{1}{t^{15}_{\Omega}} \int_{0}^{t^{15}_{\Omega}} u_{n}^{\ast}(\tilde{x},\tilde{y},t + n t^{15}_{\Omega}) \varepsilon_{n}^{\ast}(\tilde{s},t+n t^{15}_{\Omega} + \tau) \mathrm{d}t,
\label{eq:correlation_starstar}
\end{multline}

where $t^{15}_{\Omega}$ is the period of 15 combined rotor cycles and $N_{15,\mathrm{cycles}}$ is the number of such intervals within the total acquisition time $T$. This extended averaging mitigates cycle-to-cycle variability while preserving the temporal structure of wake-blade coupling over longer time scales.

\begin{figure*}
  \includegraphics[width=\textwidth]{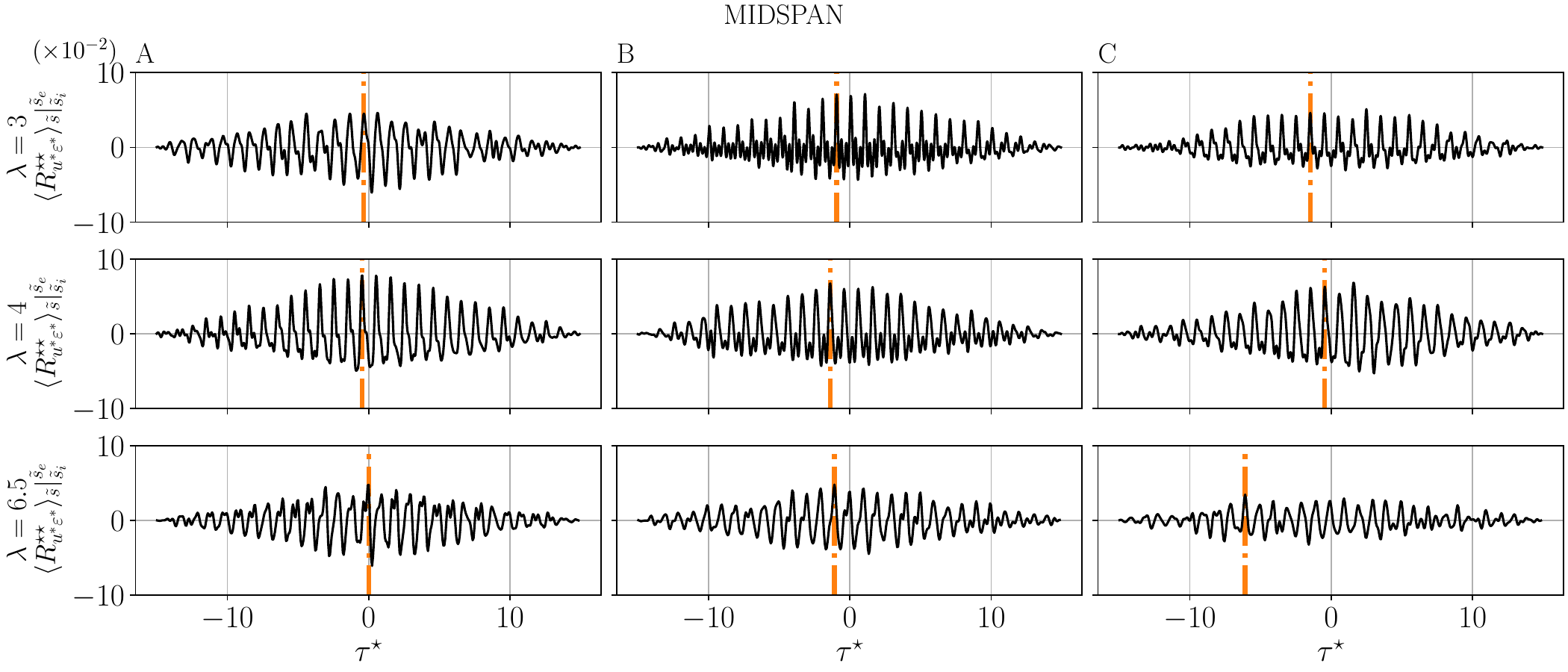}
\caption{Extended cycle-averaged cross-correlation 
  $R_{u^{\ast}\varepsilon^{\ast}}^{\star\star}$ over 15 rotor cycles 
  between wake velocity fluctuations at $\tilde{y}=0.5$, $\tilde{x}=0.7$ 
  and blade strain at the $\mathrm{MIDSPAN}$, as a function of $\tau^{\star}$, 
  $\lambda$ and FST condition.}
  \label{fig:CROSSCORRZOOM}
\end{figure*}

We focus on the blade section and wake region exhibiting the strongest coupling in figure \ref{fig:fig13new}: the $\mathrm{MIDSPAN}$ and the shear-layer region at $\tilde{y}=0.5$. Figure \ref{fig:CROSSCORRZOOM} presents $R_{u^{\ast}\varepsilon^{\ast}}^{\star\star}$ as a function of normalized time lag $\tau^{\ast}$ for all FST conditions and $\lambda$ tested.

The extended cycle-averaged cross-correlations exhibit an oscillatory structure across the $15$-cycle window, with envelope modulation that depends on both $\lambda$ and FST condition. 
Consistent with figure \ref{fig:fig13new}, figure \ref{fig:CROSSCORRZOOM} shows that increasing $\mathrm{TI}$ systematically reduces the correlation magnitude, while $\lambda=4$ shows the largest magnitude, and most sustained correlation magnitude across all FST conditions. 
For $\lambda=6.5$, the magnitude of the extended cycle-averaged cross-correlation is substantially smaller than the remaining test cases as seen before.

The oscillatory envelope of $R^{\star\star}_{u^{\ast}\varepsilon^{\ast}}$ exhibits a predominantly negative-lag peak across operating conditions and FST levels, suggesting that strain fluctuations tend to precede the corresponding velocity fluctuations detected at $\tilde{x}=0.7$. This is physically consistent with blade-generated forcing, whereby flow structures originating at the rotor plane, interact with the rotor's blades, and subsequently convect downstream to the measurement station. Such directionality is consistent with observations by \citet{Yadala2025JFM} on elastic airfoils, where structural motion was similarly found to precede downstream wake perturbations, though in the present rotating configuration this coupling is further organised by rotation-coherent frequencies.

However, the relatively flat envelope of $R^{\star\star}_{u^{\ast}\varepsilon^{\ast}}$---particularly at intermediate $\lambda$ and higher $\mathrm{TI}$---and the inherent causality bias discussed in section \S~\ref{sec:causality} preclude robust quantitative identification of a characteristic lag. The observed negative-lag tendency should therefore be interpreted as qualitative evidence of blade-to-wake directionality rather than a precise convective delay. Measurements closer to the rotor plane, such as high-speed PIV at $\tilde{x} < 0.5$, would be needed to establish this relationship quantitatively.

\section{Cross-correlation spectral analysis}

To resolve the frequency-dependent coupling between wake dynamics and blade response, we compute the cross-power spectral density ($\mathrm{CPSD}$) between the spanwise strain fluctuations, $\varepsilon$, and streamwise wake velocity fluctuations, $u$. The $\mathrm{CPSD}$ is obtained via Fourier transform of the cross-correlation function:
\begin{equation}
\mathrm{CPSD}(\tilde{x},\tilde{y},\tilde{s},St_{\Omega}) =\mathcal{F}_{\tau \rightarrow St_{\Omega}}\left[ R_{u\varepsilon}(\tilde{x},\tilde{y},\tilde{s},\tau) \right],
\label{eq:CPSD}
\end{equation}
where $St_{\Omega} = f/F_R$ is the Strouhal number normalized by the rotor frequency. 

As a complex-valued function, the $\mathrm{CPSD}$ encodes both coupling strength (magnitude  $\lvert \mathrm{CPSD} \rvert$) and relative phase (argument  $\angle (\mathrm{CPSD}$)) between the flow and structural signals. We focus on $\lvert \mathrm{CPSD} \rvert$ to quantify frequency-resolved correlation strength, as the phase $\angle (\mathrm{CPSD}$) is likely to be compromised by the causal bias discussed so far inherent to the experimental methodology.

\begin{figure*}
  \centering
  \raisebox{2.8in}{\textit{a)}}\includegraphics[width=\textwidth]{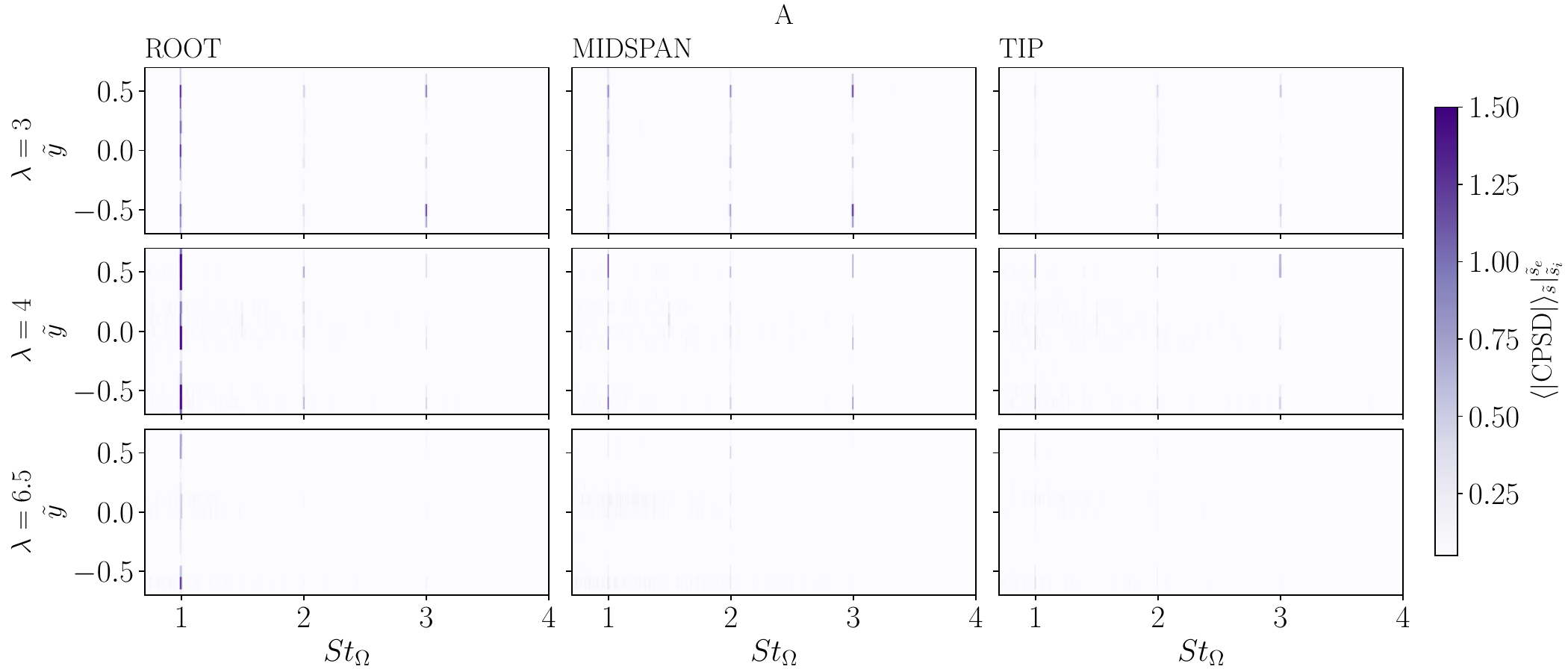}
  \raisebox{2.8in}{\textit{b)}}\includegraphics[width=\textwidth]{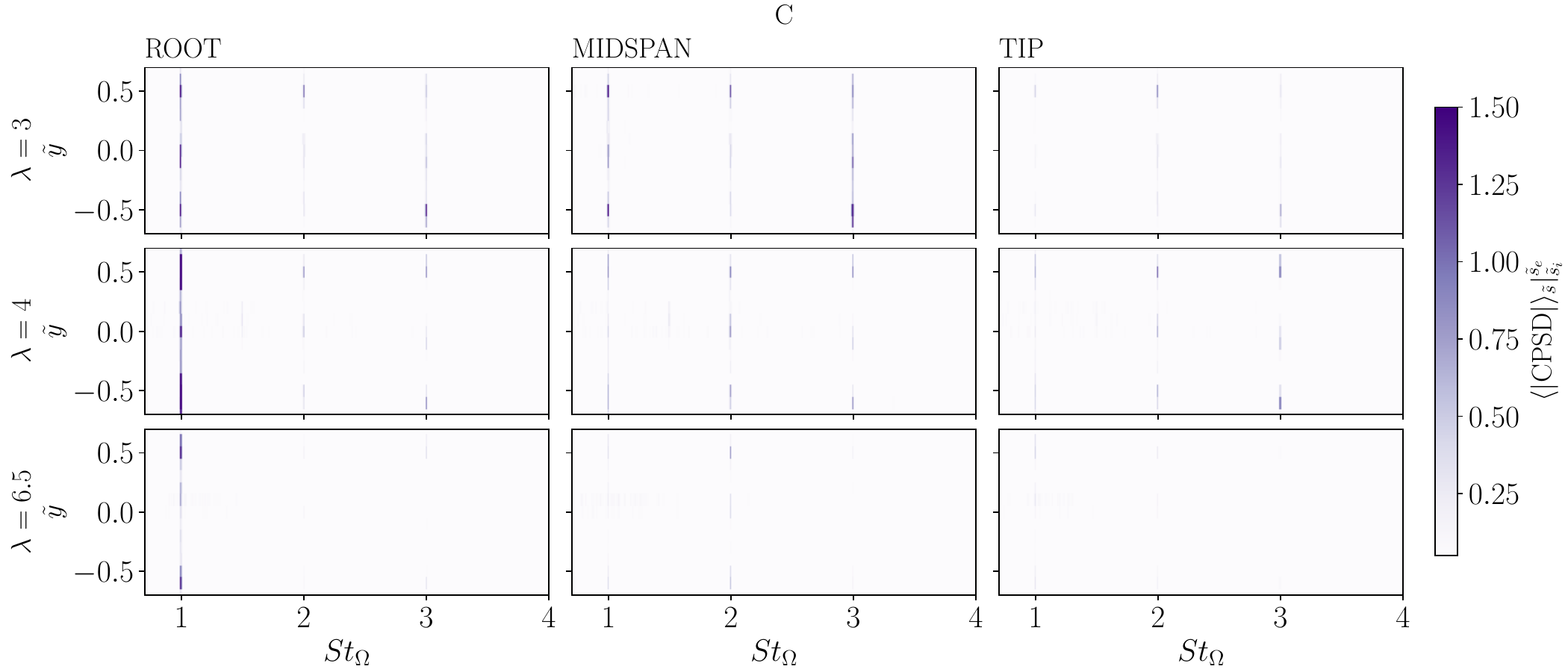}
  \caption{$\langle\vert \mathrm{CPSD}\vert \rangle_{\tilde{s}}\vert_{\tilde{s}_i}^{\tilde{s}_e}$ maps for the three sections of the blade under analysis, $\lambda\in\{3,4,6.5\}$ and the FST conditions $\mathrm{A}$ (\textit{a)}) and $\mathrm{C}$ (\textit{b)}), at $\tilde{x} = 0.7$ across $\tilde{y}$. Used with permissison from \citet{francisco3}.}
  \label{fig:CPSD2Dmaps}
\end{figure*}

Figure \ref{fig:CPSD2Dmaps} presents the spanwise-averaged $\lvert \mathrm{CPSD}\rvert$ at $\tilde{x} = 0.7$ for each blade section (averaged between measurement stations $\tilde{s}_i$ and $\tilde{s}_e$), shown as a function of wake transverse position $\tilde{y}$ and $St_\Omega$ for representative $\lambda$ and FST conditions $\mathrm{A}$ and $\mathrm{C}$.

Figure \ref{fig:CPSD2Dmaps} shows that wake-blade coupling at $\tilde{x} = 0.7$ is organized around discrete harmonics of the rotor frequency, with dominant contributions at $St_{\Omega}\in \{1,2,3\}$ spatially localised within the wake shear layers ($\lvert\tilde{y}\rvert \approx 0.4$-$0.6$). 
The coupling strength decreases progressively from $\mathrm{ROOT}$ to $\mathrm{TIP}$, reflecting the spanwise variation in induced bending moment by the various dynamics acting across the extent of the blade (see \citep{francisco3} for more details). 

\begin{figure*}
  \centering
  \raisebox{3.2in}{\textit{a)}}\includegraphics[width=\textwidth]{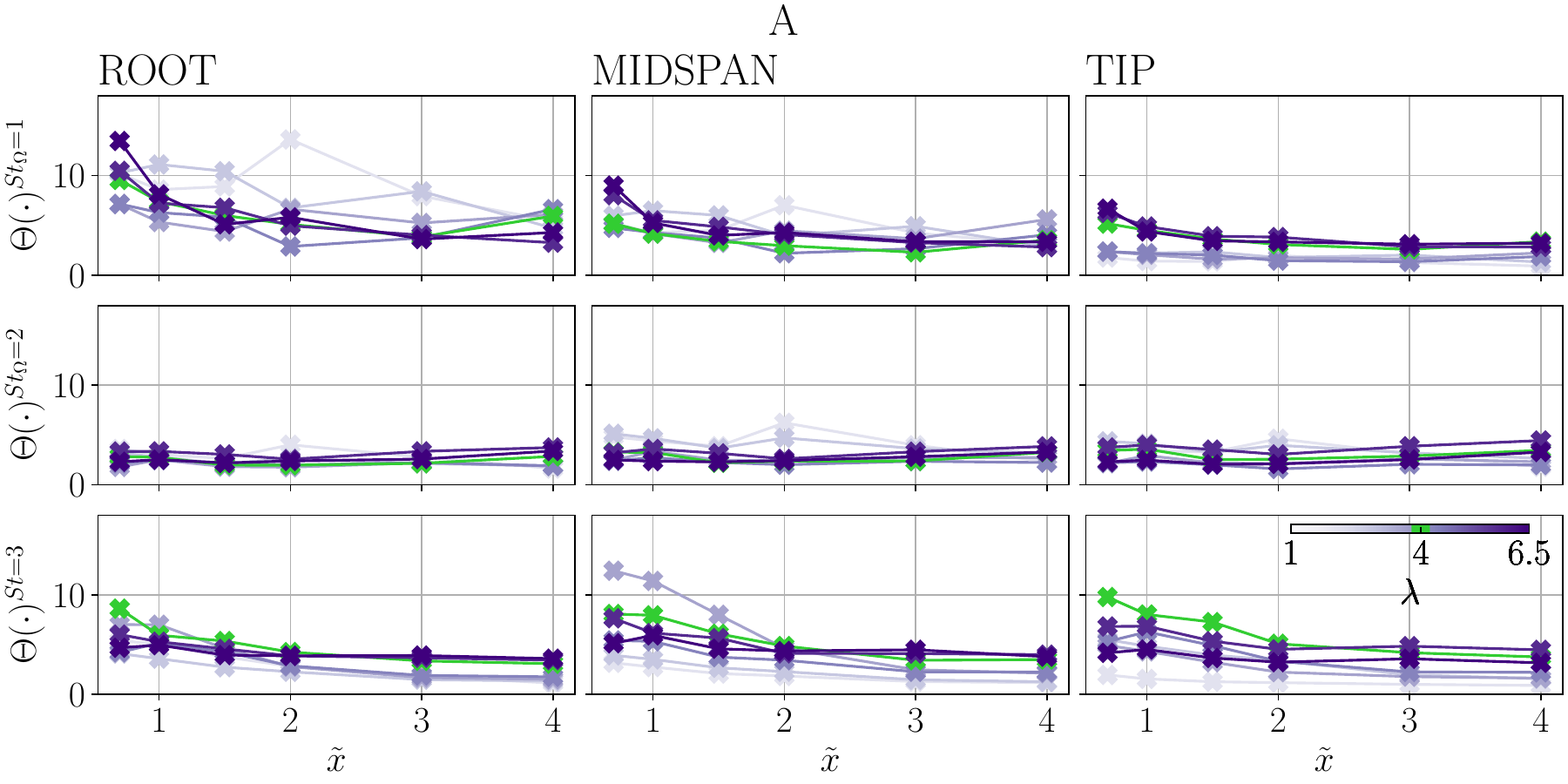}
  \raisebox{3.2in}{\textit{b)}}\includegraphics[width=\textwidth]{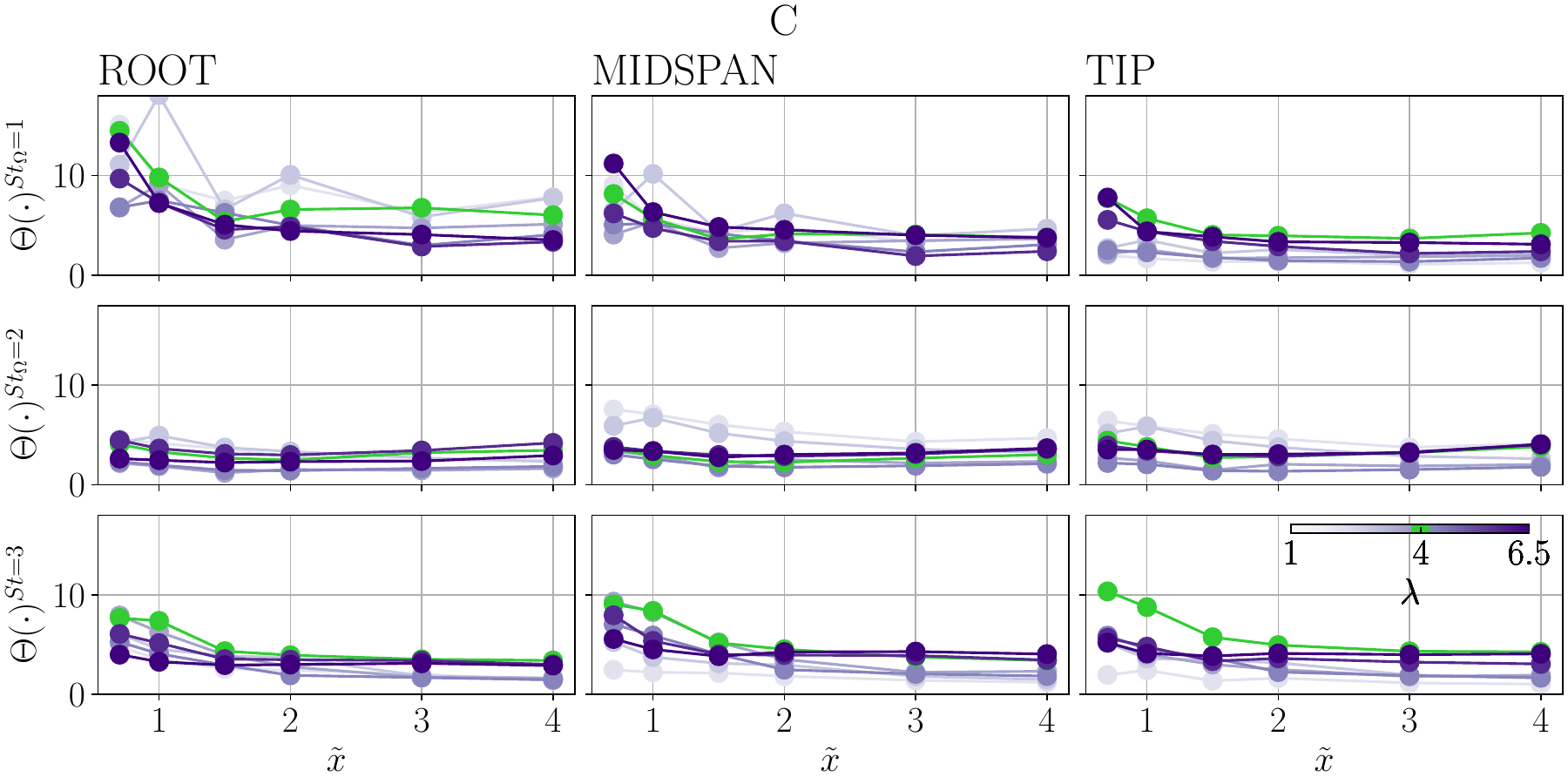}
\caption{Energy integration of $\vert \mathrm{CPSD}\vert$ between $u$ and $\varepsilon$ at $St_{\Omega}\in\{1,2,3\}$ across $\lambda$, $\tilde{x}$, and FST.}
  \label{fig10}
\end{figure*}

To quantify the relative importance of each relevant dynamic at $St_{\Omega}\in\{1,2,3\}$ and allow the comparison across operating conditions, we integrate $\lvert\mathrm{CPSD}\rvert$ over frequency bands centred on $St_\Omega \in \{1,2,3\}$, accounting for the spatial extent of coupling across both blade and wake measurement locations. The strain dynamics over the $3$ blade regions $[\mathrm{ROOT, MIDSPAN, TIP}]$ under analysis are respectively correlated to the velocity fluctuations in spanwise stations $\tilde{y} \in \{[-0.1,0.1], \pm[0.3,0.2], \pm[0.6,0.4]\}$, maintaining the spatial structure of the regions of the flow to the respective portions on the blade. $\Theta(\vert\mathrm{CPSD}\vert)$ is then defined as:
\begin{equation}
\Theta(\vert\mathrm{CPSD}\vert)(\tilde{x}) = \sum_{s=\tilde{s}_i}^{\tilde{s}_e} \sum_{y={\tilde{y}_i}}^{\tilde{y}_e} \int_{St^\prime_{\Omega}-W/2}^{St^\prime_{\Omega}+W/2} \Gamma \quad\mathrm{d}St'_{\Omega},
\end{equation}
centred on $St_{\Omega}^{\prime}$ for $St_{\Omega}\in\{1,2,3\}$ where $\{\tilde{s}_i,\tilde{s}_e\}$ and $\{\tilde{y}_i,\tilde{y}_e\}$ correspond to the spatial limits of the regions of the flow and blade to be correlated, $W$ to the window used for the integration (we select $W=0.2$ having conducted a sensitivity study) and $\Gamma$ to $\vert\mathrm{CPSD}(\tilde{x},\tilde{y},\tilde{s},St\prime_{\Omega})\vert$. $\Theta$ quantifies the correlated energy between the strain and velocity signals, not the direct aerodynamic work acting on the blade. Figure \ref{fig10} presents $\Theta(\vert\mathrm{CPSD}\vert)$ for $St_{\Omega}\in\{1,2,3\}$ over the $\mathrm{ROOT}$, $\mathrm{MIDSPAN}$ and $\mathrm{TIP}$, for different $\lambda$ and FST cases $\mathrm{A}$ (figure \ref{fig10} \textit{a)}) and $\mathrm{C}$ (figure \ref{fig10} \textit{b)}), along the flow streamwise extent $\tilde{x}$. 

We begin the analysis of figure \ref{fig10} by examining the influence of flow structures at $St_{\Omega} = 3$, which are linked to tip-vortex formation in the turbine wake's shear layer, on the evolution of the $\mathrm{CPSDs}$. The streamwise decay of $\Theta(\vert\mathrm{CPSD}\vert)^{St_\Omega=3}$ with increasing $\tilde{x}$ indicates the gradual dissipation of tip vortices formed in the near-wake region, resulting in a reduced contribution to cross-correlation. Across all tested $\lambda$, $\Theta(\vert\mathrm{CPSD}\vert)^{St_{\Omega} = 3}$ consistently peaks at $\tilde{x}=0.7$, with subtle section-to-section variations reflecting how tip-generated flow structures propagate and influence different blade regions.

As $\lambda$ increases beyond $4$, the differences in the magnitudes of $\Theta(\vert\mathrm{CPSD}\vert)^{St_{\Omega} = 3}$ between blade sections become less pronounced. Additionally, higher $\mathrm{TI}$ in the FST leads to a relative increase in $\Theta(\vert\mathrm{CPSD}\vert)^{St_{\Omega} = 3}$ across all tested $\lambda$ and blade sections. This effect is particularly notable at $\lambda \approx 4$ and at the $\mathrm{MIDSPAN}$ and $\mathrm{TIP}$, suggesting that, at design $\lambda$, increased FST $\mathrm{TI}$ enhances the energy of the coupled dynamics between wake tip vortices and the blade. While the $\mathrm{TIP}$ is where these flow structures originate, the $\mathrm{MIDSPAN}$ appears especially sensitive due to the accumulation of bending effects. 

\begin{figure}[h!]
  \centering
  \includegraphics[width=\columnwidth]{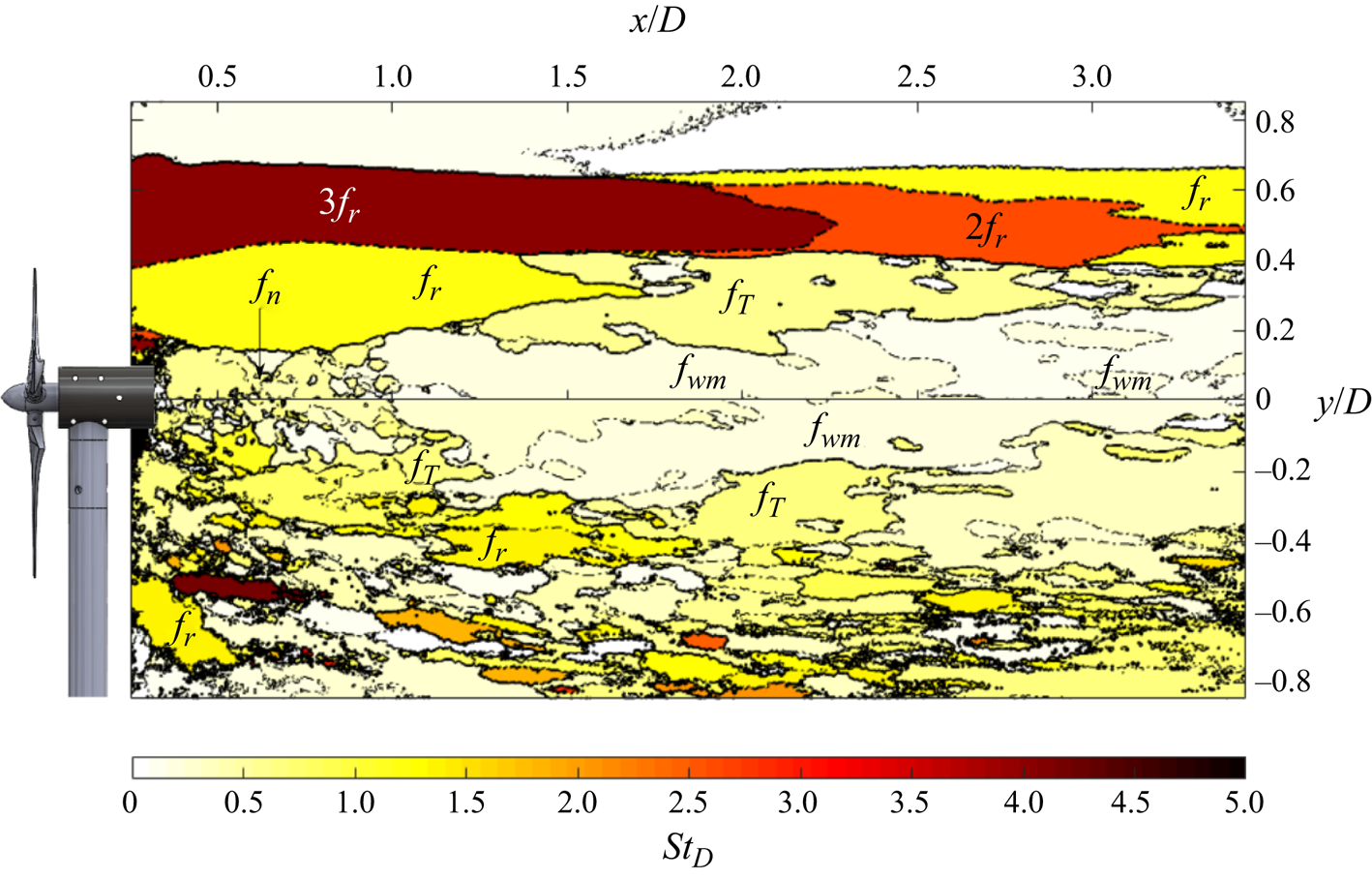}
  \caption{Zones of dominant frequencies in a wind turbine's wake operating at $\lambda=4.5$ across the wind turbine's spanwise ($z/D$) and streamwise extent ($x/D$). Map entails the multitude of dynamics to which a wind turbine's rotor is exposed. $f_T$, $f_n$, $f_{wm}$ and $f_r$ correspond respectively to tower shedding, nacelle shedding, wake meandering and rotor characteristic frequencies. $St_D$ corresponds to the Strouhal number normalised with the wind turbine's diameter ($St_D = f \times D / U_{\infty}$). Reproduced with permission from \citep{biswas2024a}, figure 9 (a) from the original paper.}
  \label{fig:neelakash}
\end{figure}

The evolution of $\Theta(\vert\mathrm{CPSD}\vert)^{St_{\Omega} = 1}$ consistently decays with $\tilde{x}$ for all three blade sections and $\lambda$ values considered apart from $\lambda=1$ where it peaks at $\tilde{x}=2$, at the $\mathrm{ROOT}$, with the greatest energy reduction occurring in the highest turbulence case ($\mathrm{C}$). 
In contrast, $\Theta(\vert\mathrm{CPSD}\vert)^{St_{\Omega} = 2}$ follows a non-monotonic pattern with $\tilde{x}$ across $\lambda$ and FST cases, initially decreasing after the rotor plane.
$St_{\Omega}=2$ motions arise from triadic interactions between $St_{\Omega}=1$ and $St_{\Omega}=3$ dynamics, with their peak energy content located further downstream of the rotor plane than that of the $St_{\Omega}=3$ motions, reflecting the streamwise development of the triadic energy transfer \citep{biswas2024a}. 
Figure \ref{fig:neelakash} is a reproduction from the results obtained in \cite{biswas2024a} of the dominant flow frequencies across a wind turbine's wake at hub height operating at $\lambda=4.5$ detailing these mechanisms. As the wake develops, at a streamwise distance from the rotor plane $St_{\Omega}=2$ becomes the dominant flow dynamic as $St_{\Omega}=3$ and $St_{\Omega}=1$ interact. The multitude of frequencies across the rotor's span also highlights the large ensemble of dynamics to which a wind turbine's blade is exposed. 

Coming back to the analysis of figure \ref{fig10}, the non-linear energy-decay behaviour is consistent with the wake's spectral dynamics documented by \citet{biswas2024a}, who showed that for their turbine and $\lambda \in \{4.5, 6\}$, the energy associated with $St_{\Omega}=3$ decays monotonically downstream of the rotor, while $St_{\Omega}=2$ exhibits local maxima at $\tilde{x} \in \{0.5, 1.5\}$ and $St_{\Omega}=1$ peaks near $\tilde{x} \in \{0.7, 1.5\}$, reflecting the complex redistribution of energy across wake dynamics through nonlinear interactions. 
This highlights the streamwise and operation-dependent inhomogeneity of flow dynamics populating the turbine's wake, with direct and indirect effects on the blade response depending on the causal relationship between wake evolution and blade forcing.

\begin{figure*}[t]
  \centering
  \includegraphics[width=\textwidth]{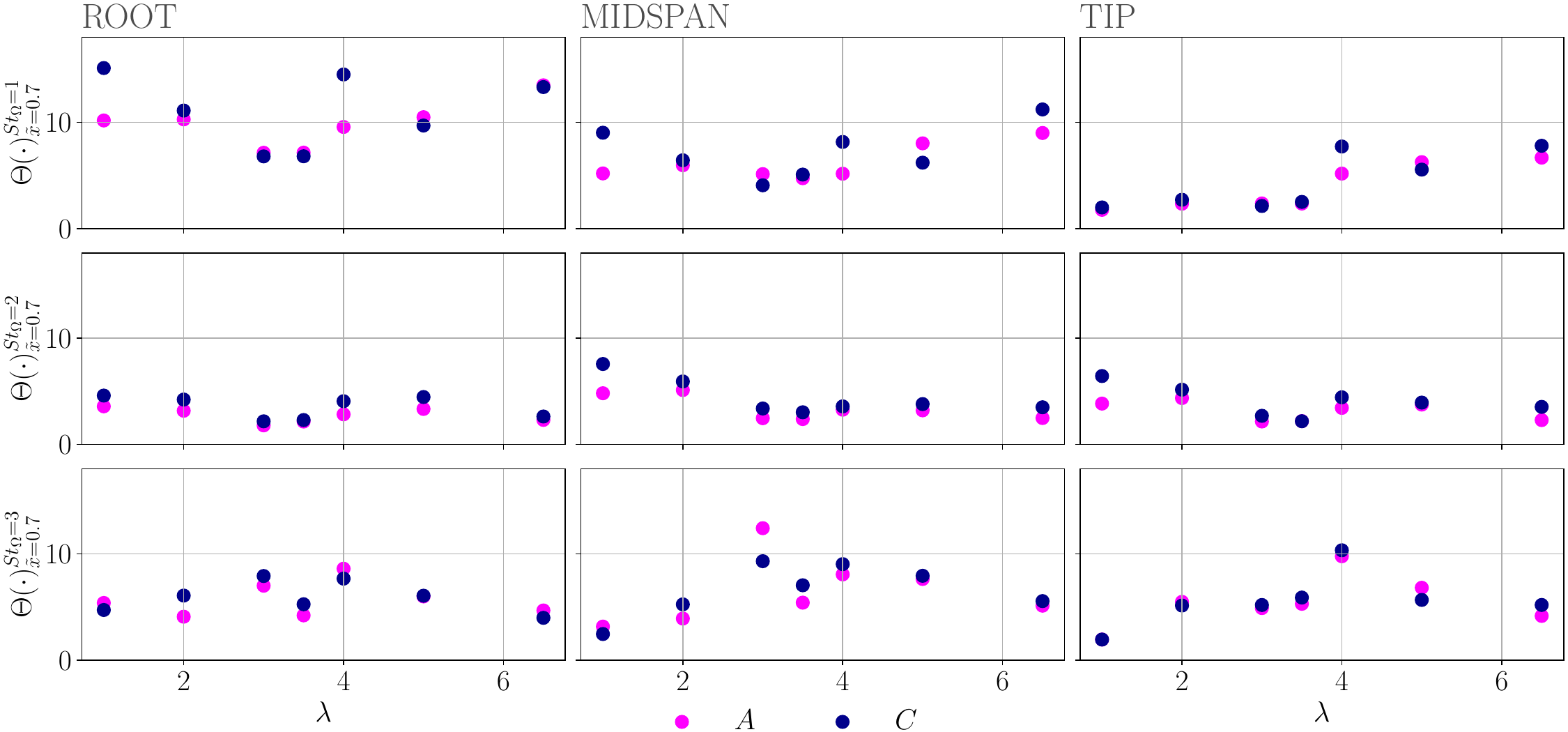}
  \caption{Distribution of $\Theta(\vert\mathrm{CPSD}\vert)^{St_{\Omega}\in\{1,2,3\}}$ at $\tilde{x}=0.7$ as a function of $\lambda$ and FST.}
  \label{fig11}
\end{figure*}

While the largest energy for $St_\Omega = 1$ and $St_\Omega = 3$ is generally observed near $\tilde{x} = 0.7$, the $St_\Omega = 2$ band shows weaker streamwise variation, with a $\lambda$-dependent trend: below-design $\lambda$ cases peak closer to the rotor, while for above-design $\lambda$ cases grow modestly with $\tilde{x}$ as the flow structure develops through triadic interactions between $St_\Omega = 1$ and $St_\Omega = 3$ dynamics \citep{biswas2024a}. 
Considering the causal interpretation of the negative-lag observed between the two signals considered in this work, on average, blade deflections lead the flow structures that are advected to the hot-wire probes. For $St_\Omega = 2$, however, these flow structures are not formed until further downstream, or farther than the immediate region of the rotor plane, where they rely on the $St_\Omega = 1$--$St_\Omega = 3$ interaction. The corresponding blade signature on $St_\Omega = 2$ is therefore more plausibly an imprinting of these motions on the blade dynamics through pressure propagation. The wake is then expected to lead the blade for these specific dynamics.
To further interpret these findings, figure \ref{fig11} complements figure \ref{fig10} by detailing the distribution of $\Theta(\vert\mathrm{CPSD}\vert)^{St_{\Omega}\in\{1,2,3\}}$ for the different combinations of $\lambda$ and FST for the $3$ blade sections at $\tilde{x}=0.7$.

From figure \ref{fig11}, $\Theta(\vert\mathrm{CPSD}\vert)^{St_{\Omega} =1}$ peaks at the $\mathrm{ROOT}$ for all $\lambda$ and FST conditions. Moreover, $\Theta(\vert\mathrm{CPSD}\vert)^{St_{\Omega} =1}$ consistently exhibits the lowest magnitude at the $\mathrm{TIP}$ among the assessed blade sections, reflecting differences in the interaction of generated flow structures across the blade. 
This is in line with the results \citep{biswas2024a}, with the respective results depicted in figure \ref{fig:neelakash}, where $St_{\Omega}=1$ dynamics have been observed to be more energetic close to the root of the blades, whereas $St_{\Omega}=3$ related dynamics are more energetic close to the tip \citep{biswas2024a}. 
FST modulates once more the energy distribution of $\Theta(\vert\mathrm{CPSD}\vert)^{St_{\Omega} =1}$ over the tested range of $\lambda$. For each $\lambda$ and blade section, increased $\mathrm{TI}$ in the FST results in a higher $\Theta(\vert\mathrm{CPSD}\vert)^{St_{\Omega} =1}$. 
In addition to modulating the amplitude of $\Theta(\vert\mathrm{CPSD}\vert)^{St_{\Omega} =1}$, increased FST intensity significantly alters the location of the peak of $\Theta(\vert\mathrm{CPSD}\vert)^{St_{\Omega} =1}$ along $\tilde{x}$, likely due to the effect that FST has on the evolution/dissipation of flow structures within the wake. 

The distribution of $\Theta(\vert\mathrm{CPSD}\vert)^{St_{\Omega} =2}$ remains relatively homogeneous across the blade regions at a fixed $\lambda$ for $\lambda\geq4$ (see figure \ref{fig11}). As $\lambda$ increases beyond $\lambda=3.5$, $\Theta(\vert\mathrm{CPSD}\vert)^{St_{\Omega} =2}$ increases in magnitude and remains relatively constant across the tested cases. The introduction of larger $\mathrm{TI}$ in the FST increases the magnitude of $\Theta(\vert\mathrm{CPSD}\vert)^{St_{\Omega} =2}$, especially for $\lambda<4$. Moreover, this effect is most pronounced at the $\mathrm{MIDSPAN}$. These results suggest that for $St_{\Omega}\in\{1,2\}$, the correlations between structural and flow dynamics are most sensitive to the presence of FST. Meanwhile, the $\mathrm{TIP}$ correlation to wake dynamics is sensitive to $St_{\Omega} \in \{2,3\}$ and less sensitive to the presence of increased $\mathrm{TI}$ in the FST.

\section{Conclusion}

This study has presented an experimental investigation of wake-blade coupling in a model wind turbine using concurrent, spatially and temporally resolved measurements of wake and blade dynamics. By systematically varying the tip-speed ratio, $\lambda$, whilst exposing the wind turbine to different FST conditions, we produced a large ensemble of datasets representative of realistic operating environments. The blade's strain response was found to be primarily governed by $\lambda$ building on our previous work \citep{francisco3}, which controls the amplitude, coherence, and spectral organisation of the structural dynamics along the blade's span. Free-stream turbulence, by contrast, predominantly modulates the magnitude of these dynamics without fundamentally altering their underlying spatial contribution to the different sections of the blade.
In our case, hot-wire anemometry doesn't allow us to attain velocity measurements as close to the blade as $2$D-PIV would allow. However, this creates a foundation to analyse the dynamic evolution of the coupling between a blade of a wind-turbine and its wake at a finer scale than the one assessed here, such that potential non-linear behaviour is well captured.

Wake-blade coupling is shown to be spatially localised within the wake shear layers, and organised around rotation-coherent frequencies, indicating that blade dynamics are linked primarily to structured, rotor-synchronous flow features rather than to wake-core turbulence. The non-monotonic streamwise evolution of coupling strength further demonstrates that wake-induced loading is maximised after partial wake development, where coherent structures interact with broadband fluctuations prior to their breakdown. Operating conditions associated with a large induced fluctuating structural response do not necessarily coincide with strong wake correlation, as evidenced by $\lambda=3.5$, highlighting the need to distinguish blade-local aerodynamic phenomena from wake-mediated forcing.

The wake-blade interaction identified here is therefore not instantaneous, but emerges through lagged, frequency-specific coupling mechanisms. The correlation-based analysis does not seek to establish direct aerodynamic causality between individual wake structures and blade response; rather, it quantifies statistically repeatable, rotor-synchronous coupling that reflects wake-mediated signatures of blade-exciting dynamics. Within this framework, the predominantly negative-lag character of the cross-correlations qualitatively indicates that blade strain fluctuations tend to precede the corresponding wake velocity fluctuations, consistent with blade-generated structures immediately after the rotor plane convecting downstream to the measurement station. Even though this directionality is not established quantitatively, it is robust across operating conditions and FST levels. 
This directionality has been assessed in a time-averaged sense, without separating the contributions of the different dynamics evolving in the turbine's wake. Resolving these contributions would likely reveal how flow structures associated with $St_{\Omega}=2$ are likely imprinted from the wake into the blade, as these form through triadic interactions between $St_{\Omega}=3$ and $St_{\Omega}=1$ typically further downstream in the wake \citep{biswas2024a}.

These findings underscore the importance of accounting for operating conditions when assessing wake-induced fatigue and structural dynamics in wind turbines. Metrics based solely on strain amplitude or coherence may misrepresent the contribution of wake dynamics if the underlying coupling mechanisms are not properly both spatially and temporally resolved. The concurrent, spatially resolved flow-structure framework presented here provides a pathway to isolate fatigue-relevant, wake-mediated loading from local aerodynamic effects, with direct implications for reduced-order modelling, digital-twin development, and condition-based monitoring strategies.

\acknowledgements{F. J. G. de Oliveira would like to thank fruitful discussions with Adrian T. McGlade in the conception of this paper. We acknowledge the help provided by Ricardo Huerta Cruz, Paul Howard and Kevin Gouder during set-up. We also acknowledge Roland Hutchins, Franco Giammaria, Mark Grant for advice and help with the production of the wind turbine model, and its respective control system. F. J. G. de Oliveira and O. R. H. Buxton wish to acknowledge financial support given by EPSRC through grant no. EP/V006436/1. Use was also made of EPSRC grant no. EP/L024888/1 for access to the National Wind Tunnel Facility (NWTF).}

\bibliography{references}

\end{document}